\documentclass[paper]{JHEP}
\usepackage{pictex,rotating,amssymb}
\usepackage{epsfig}

\newcommand{\newc}{\newcommand}
\newc\eg{{\it {e.g.}}}  \newc\etal{{\it {et al.}}} \newc\ie{{\it i.e.}}

\newc\second{{\rm sec}} 

\newc\mone{M_1} % plasmon mass

\newc{\logsoms}{\log\left(s/m_{\rm eff}^2\right)}

\newc\alphas{\alpha_s}

\newc{\gstar}{g_\ast}           \newc{\gsstar}{g_{s\ast}}
\newc{\geff}{g_{\rm eff}}

\newcommand\mz{m_{Z}}
\newc\meff{m_{\rm eff}} % plasmon mass

\newc\sigmabar{{\overline{\sigma}}}

\newcommand\treh{T_{\rm R}}
\newcommand\Tdec{T_{\rm dec}}   \newcommand\tdec{t_{\rm dec}}
\newcommand\Td{T_{\rm D}}
\newcommand\Teq{T_{\rm eq}}     % RD=MD equil.
\newcommand\Tnr{T_{\rm NR}}
\newcommand\tf{T_{\rm f}}       % T_freezeout
\newc{\xf}{x_f}

\newc{\nspin}{n_{\rm spin}}
\newc{\nflavor}{n_{\rm F}}

\newc\vrel{v_{\rm rel}}
\newcommand\fa{f_{a}}
\newcommand\mchi{m_{\chi}}              \newcommand\nchi{n_{\chi}}
\newcommand\tauchi{\tau_{\chi}}

\newcommand\squark{\widetilde q}        \newcommand\msquark{m_{\squark}}

\newcommand\gluino{\widetilde g}
\newcommand\mgluino{m_{\widetilde g}}   \newcommand\ngluino{n_{\widetilde g}}

\newcommand\axino{\widetilde{a}}        \newcommand\maxino{m_{\axino}}
\newcommand\saxion{s}        \newcommand\msaxion{m_{\saxion}}

\newcommand\gravitino{\widetilde{G}}    \newcommand\mgravitino{m_{\gravitino}}

\newcommand\mlosp{m_{\rm LOSP}}

\newcommand\abunda{\Omega_{\axino}h^2}  \newcommand\omegaa{\Omega_{\axino}}

                \newcommand\omegaantp{\Omega^{\rm NTP}_{\axino}}
       
                \newcommand\omegaatp{\Omega^{\rm TP}_{\axino}}

\newc{\cachigamma}{C_{a\chi\gamma}}     \newc{\cachigammai}{C_{a\chi\gamma,i}}

\newc{\caww}{C_{aWW}}                   \newc{\cayy}{C_{aYY}}
\newc{\cagg}{C_{agg}}
\newc{\sthw}{\sin\theta_W}              \newc{\cthw}{\cos\theta_W}
\newc{\bino}{\widetilde B}              \newc{\wino}{\widetilde W_3}
\newc{\higgsinob}{{\widetilde H}^0_b}   \newc{\higgsinot}{{\widetilde H}^0_t}

\newc{\naxino}{n_{\tilde a}}
\newc{\ngamma}{n_\gamma}

\newc{\ychi}{Y_{\chi}}                  \newc{\yeqchi}{Y^{\rm EQ}_{\chi}}

\newc{\yaxino}{Y_{\axino}}
\newc{\yeqaxino}{Y^{\rm EQ}_{\axino}}
\newc{\ythaxino}{Y^{\rm TP}_{\axino}}
\newc{\ynthaxino}{Y^{\rm NTP}_{\axino}}
\newc{\yascat}{Y^{\rm scat}_{i,j}}      \newc{\yadec}{Y^{\rm dec}_{i}}

\newc{\abund}{\Omega h^2}
\newc{\abundchi}{\Omega_\chi h^2}
\newc{\rhocrit}{\rho_{\rm crit}}
\newc{\rhochi}{\rho_{\chi}}
\newc{\rhoaxino}{\rho_{\axino}}
\newc{\rhonu}{\rho_{\nu}}

\newc{\deltannu}{\delta N_{\nu}}

\newc{\mplanck}{M_{\rm P}}      \newc{\mpl}{M_{\rm Pl}}
\newc{\msusy}{M_{\rm SUSY}}      \newc{\ms}{M_{\rm S}}

\newc{\jxf}{J({\xf})}
\newc{\VEV}[1]{\langle #1 \rangle}

\newcommand\tev{\,\mbox{TeV}}
\newcommand\gev{\,\mbox{GeV}}
\newcommand\mev{\,\mbox{MeV}}
\newcommand\kev{\,\mbox{keV}}
\newcommand\ev{\,\mbox{eV}}
\newc{\ra}{\rightarrow}
\newc{\beq}{\begin{equation}}
\newc{\eeq}{\end{equation}}
\newc{\bea}{\begin{eqnarray}}
\newc{\eea}{\end{eqnarray}}
\renewcommand\({\left(}
\renewcommand\){\right)}
\renewcommand\[{\left[}
\renewcommand\]{\right]}

\newcommand\lsim{\mathrel{\rlap{\lower4pt\hbox{\hskip1pt$\sim$}}
    \raise1pt\hbox{$<$}}}
\newcommand\gsim{\mathrel{\rlap{\lower4pt\hbox{\hskip1pt$\sim$}}
    \raise1pt\hbox{$>$}}}

\preprint{CERN-TH/2000-378, DESY 00-193, SNUTP 00-034, hep-ph/0101009}
\title{Axinos as Dark Matter }
\author{Laura Covi \\
    DESY Theory Group, Notkestrasse 85, 22603 Hamburg, Germany}
\author{Hang Bae Kim \\
    Department of Physics, Lancaster University, Lancaster LA1 4YB, England}
\author{Jihn E. Kim \\
    Department of Physics and Center for Theoretical Physics,\\
    Seoul National University, Seoul 151-747, South Korea}
\author{Leszek Roszkowski \\
    Theory Division, CERN, CH-1211, Geneva 23, Switzerland, and\\
    Department of Physics, Lancaster University, Lancaster LA1 4YB, England}
\abstract{
Supersymmetric extensions of the Standard Model that incorporate
the axion solution to the strong CP problem necessarily contain
also the axino, the fermionic partner of the axion. In contrast to
the neutralino and the gravitino, the axino mass is generically not
of the order of the supersymmetry-breaking scale and can be much
smaller. The axino is therefore an intriguing candidate for a stable
superpartner. In a previous Letter~[1] %\protect\cite{ckr}
it was shown that
axinos are a natural candidate for cold dark matter in the Universe
when they are generated non-thermally through out-of-equilibrium
neutralino decays.  Here, we extend the study of non-thermal
production and include a competing thermal production mechanism
through scatterings and decays of particles in the plasma. We
identify axino masses in the range of tens of~MeV to several~GeV
(depending on the scenario) as corresponding to cold axino relics if
the reheating temperature $\treh$ is less than about
$5\times10^4\gev$. At higher $\treh$ and lower mass, axinos could
constitute warm dark matter. In the scenario with axinos as relics,
the gravitino problem finds a natural solution. The lightest
superpartner of the Standard Model spectrum remains effectively stable in
high-energy detectors but may be either neutral or charged. The
usual highly
restrictive constraint $\abundchi\lsim1$ on the relic abundance of the
lightest neutralino becomes void.
}
\begin{document}

%%%%%%%%%%%%%%%%%%%%%%%%%%%%%%%%%%%%%%%%%%%%%%%%%%%%%%%%%%%%%%%%%%%%%%%%%%%
\section{Introduction}
\label{sec:introduction}

Among several possible ways of addressing theoretical puzzles of the
Standard Model, two are generally believed, for quite independent
reasons, to be very compelling.  Softly-broken low-energy
supersymmetry (SUSY) seems to be the most attractive way of solving
the hierarchy problem and of linking the electroweak scale with
physics around the Planck scale. Moreover, it has the generic property
of contributing only small corrections to electroweak parameters and
also of predicting a light Higgs boson, both in agreement with the
outcome of current precision measurements.

The most compelling way of resolving the strong CP problem of QCD
seems to be provided by invoking the Peccei and Quinn (PQ)
mechanism~\cite{pq}. There, the CP-violating $\theta$-term in the QCD
Lagrangian, which is experimentally required to be excessively tiny,
is replaced by a term involving a new fundamental pseudo-scalar field,
the axion. This is achieved by introducing a global, chiral $U(1)$
symmetry group, which is spontaneously broken at some high energy scale
$\fa\sim10^{11}\gev$. 
The QCD anomaly breaks this chiral $U(1)$ symmetry at the one-loop level,
and hence the axion becomes not a true Goldstone boson but a
pseudo-Goldstone boson with a tiny mass of order $\Lambda_{\rm
QCD}^2/f_a$~\cite{qcdanomaly:cite}. As the lightest pseudo-scalar
consistent with the known particle phenomenology, the axion can be
very important in cosmology and astrophysical processes.
Axion physics and cosmology are very
rich fields which have been thoroughly explored in the
literature~\cite{axionreviews:cite}.

Axions can also be the dark matter (DM) in the Universe. Indeed, 
being extremely light and feebly interacting, and having a
lifetime of order $> 10^{40}$~years, axions can be
considered as stable for all practical purposes.
Therefore, if axions are copiously produced in the early Universe, they
can contribute substantially to the relic mass density. Since they are
spin-zero particles, we can consider hot axions and cold axions
separately. Hot axions are produced when the PQ symmetry is broken
at $T\le f_a$. Such hot axions are diluted by inflation if the
reheating temperature after inflation is less than about $10^9\gev$. At low
temperature it is difficult to produce  hot axions because of their
tiny interaction strength.  Since the axion potential is extremely flat,
the axion vacuum expectation value (VEV) $\langle a\rangle$
will not move until the
temperature of the Universe falls below $\sim 1\gev$. These coherent
axions produced around the QCD phase transition have very small
kinetic energy and are therefore cold. They can constitute the 
DM of the Universe
if $f_a \sim 10^{12}\gev$~\cite{axiondm}.

An inevitable prediction of combining the two well-motivated and
independent hypotheses of axions and supersymmetry is the existence of
the SUSY partner of the axion, the axino. Being massive and both
electrically and color neutral, axinos are an intriguing possibility
for the lightest supersymmetric particle (LSP) and a WIMP.  In the
presence of $R$-parity conservation there may be dramatic cosmological
implications of the existence of such stable massive particles. In
particular, they may contribute substantially to the relic mass
density and may constitute the main component of the DM in the
Universe. This is the possibility that we will explore in this study.

There are two other SUSY particles with well-known, and much-studied,
cosmological properties which, ever since the early days of low-energy
SUSY, have been considered for DM candidates. Perhaps the most popular
of them is the lightest neutralino, $\chi$, which, if stable, often
provides the desired amount of relic abundance for natural ranges of
other superpartner masses. Furthermore, since its characteristic
interaction strengths are often of a sizeable fraction of the
electroweak interactions, the neutralino has a good chance of being
detected in high energy colliders and, if it indeed constitutes the
dark matter, in WIMP dark matter searches.

The other traditional candidate, the gravitino, $\gravitino$, arises by
coupling SUSY to gravity in supergravity or superstring scenarios.  It
has been long known that gravitinos could be copiously produced in
thermal processes in the early Universe and often suffer from  the well-known
``gravitino problem''~\cite{ekn84} of excessive
destruction of light elements and of contributing too much to the
energy density.  These problems can be avoided if the reheating
temperature after inflation is sufficiently low,
$\treh\lsim10^9\gev$~\cite{moreearlygravitino,ekn84,mmy}.  Therefore,
the gravitino remains a possibility for dark matter.
(For recent work see Refs.~\cite{bbp98,ty00,wil}.)

In contrast to the neutralino and gravitino cases, axinos have,
undeservedly, attracted much less attention in the literature. This is
even more surprising in view of the fact that axinos possess some
distinctive properties that lead to important cosmological
implications. In particular, in contrast to the case of the gravitino
as well as the neutralino and other ordinary superpartners, the mass
of the axino is generically {\em not} of the order of the
SUSY-breaking scale and, as we will see later, can be much
lower~\cite{tw,rtw,ckn}. This alone makes the
axino an intriguing  candidate for the LSP and dark matter.

One important exception was provided by a comprehensive study of
Rajagopal, Turner and Wilczek~\cite{rtw}. These authors considered
both light ($\sim\kev$ range) and more massive ($\sim\gev$) axinos and
studied their cosmological implications as either the LSP or the
next-to-LSP (NLSP).  They derived an upper bound $\maxino < 2\kev$ on
the mass of stable, ``primordial'' axinos which would hold in the
absence of inflation.  They also concluded that, in a class of
interesting models, the axino could have a mass satisfying this bound
and would constitute so-called {\it `warm'} dark matter.
Such light axinos were studied also by other early papers~\cite{kmn,bgm}.

While warm DM has some interesting properties (see, \eg,
Ref.~\cite{bot}), it has been the standard cosmological lore of the
last several years that the invisible dynamical component of the
mass--energy density of the Universe is probably predominantly in the
form of {\it cold} dark matter (CDM)~\cite{kt}. The paradigm has been
based on good agreement of numerical simulations of the formation of
large scale structures with observations~\cite{kt}. More recent
studies have revealed some possible problems with predictions of the
standard CDM theory on sub-galactic scales, which still need to be
clarified.  Nevertheless, we believe that it is still well-justified
to assume that most of the DM in the Universe is cold.

In a previous paper~\cite{ckr}, we pointed out that axinos can naturally
form cold dark matter. One way of producing such CDM
axinos is through out-of-equilibrium decays of a heavy enough
superpartner.  Because axino couplings to other particles are
suppressed by $1/\fa$, as the Universe cools down, all heavier SUSY
partners first cascade-decay to the lightest ordinary SUSY particle
(LOSP), \eg, the lightest neutralino. (By ``ordinary'' we mean any of the
superpartners of the Standard Model particles.) 
The LOSPs then freeze out of thermal equilibrium and
subsequently decay into axinos. If LOSP is a neutralino of tens of
$\gev$ or heavier, its lifetime, which scales like $1/\mchi^3$, is
often much shorter than $1\, {\rm sec}$ and the decay takes place before
Big Bang Nucleosynthesis (BBN)~\cite{ckr}.

It is worth mentioning that the neutralino will nevertheless appear to be
stable in high-energy colliders since its lifetime will typically be
significantly larger than $\sim10^{-7}\, {\rm sec}$. The same would be true
if the LOSP carried electric charge. In this case the LOSP would appear in a
detector as a stable, massive, charged state~\cite{rtw}.

Furthermore, since after freeze-out all neutralinos convert into
axinos, one obtains a simple expression for the axino relic abundance
$\abunda= \left({\maxino/\mchi}\right)\, \abundchi$~\cite{ckr}. (Here
$\Omega_{\axino}=\rhoaxino/\rhocrit$, where $\rhoaxino$ is the relic density
of axinos and $\rhocrit$ is the critical density, $h$ is the
dimensionless Hubble parameter, $\maxino$ and $\mchi$ are the
respective masses of the axino and the neutralino, and $\abundchi$ is
the abundance the neutralinos would have had today.) It
follows that cases where $\abunda\sim1$ will typically correspond to
supersymmetric configurations for which $\abundchi\gg1$. In other
words, cases traditionally thought to be excluded by the condition
$\abundchi\lsim1$ will now be readily allowed (actually, even
favored!). They would also typically correspond to larger superpartner
masses, which may be unwelcome news fo collider searches, but appears to
be favored by gauge coupling unification~\cite{kkrw,fm00} as well as
flavor and proton decay constraints~\cite{dmr00,fm00}.

In addition to the non-thermal production (NTP) discussed above,
axinos can also be generated through thermal production (TP), namely
via two-body scattering and decay processes of ordinary particles
still in thermal-bath.  (We stress that, despite the name of the
process, the resulting axinos will typically be already out of thermal
equilibrium because of their exceedingly tiny couplings to ordinary
matter, except at very large $\treh$.  They will nevertheless be
produced in kinetic equilibrium with the thermal-bath and thus their
momenta will have a thermal spectrum inherited from the scattering
particles in the plasma.)  While the analogous process of gravitino
production has been studied in the literature~\cite{mmy,bbp98}, this
has not yet been done in the case of axinos.

In this study we aim to provide a detailed exploration of both TP and
NTP mechanisms of producing axinos through processes involving
ordinary superpartners, and to compare their relative effectiveness.
Both mechanisms are quite generic, but specific results will be
somewhat model-dependent. For definiteness, we will be working in the
context of the Minimal Supersymmetric Standard Model (MSSM). Also,
while any ordinary superpartner (neutral or not) could in principle be
the LOSP, we will concentrate on the case of the neutralino.  We will
also, for reference, often assume $\fa=10^{11}\gev$, although it will
be relatively straightforward to rescale our results for other values
of $\fa$. Furthermore, since the axions and axinos are produced
through different mechanisms and both can be CDM, one could consider
the possibility that both contribute sizeably to the CDM relic
abundance. While this scenario remains a viable possibility,  we will,
for definitness, explicitly assume that it is the axino that dominates
the Universe.

We will show that, unsurprisingly, TP is more important at larger
values of the reheating temperature $\treh$.  On the other hand, NTP
will dominate at rather low (but not unreasonably low) values $\treh\lsim
5\times 10^{4}\gev$. This is also, broadly, the range of $\treh$ for which
other constraints will be satisfied.

Clearly we are most interested in the cases where the relic density of
axinos $\rhoaxino$ will be close to the critical density $\rhocrit$.
More precisely, we will require $0.1\lsim\abunda\lsim0.3$.  This
condition will provide a strong constraint on the scenario. In
particular, in the NTP case of axino production through $\sim{\cal
O}(100\gev)$ bino decay, it will imply $\maxino\gsim{\cal O}(10\mev)$
to ${\cal O}(1\gev)$, depending on whether we allow the SUSY-breaking
scale $\msusy$ to take very large values (tens of $\tev$) or require
$\msusy\lsim1\tev$, respectively.

Axinos will initially be nearly always relativistic (unless 
they are nearly degenerate with the parent neutralinos), hence
they have to be heavy enough to ``cool down'' and become
non-relativistic, or cold, by the epoch of matter dominance. We find
that this leads to a somewhat weaker condition 
$\maxino\gsim{\cal O}(100\kev)$.

A similar bound arises from requiring that the relativistic axinos do
not contribute too much to the energy density of radiation during BBN,
since their production will take place near the time of BBN.
Furthermore, a significant fraction of neutralino
decays will proceed into $q\bar{q}$-pairs (+axino) through an
intermediate photon or $Z$-exchange. The resulting hadronic showers
may cause excessive destruction of light elements.  We find that this
can be easily avoided but that the specific bounds are strongly
model-dependent. It would suffice to simply assume a large enough
neutralino mass $\mchi$ for the decay to take place before BBN. We
find that such a crude requirement would be unnecessarily 
over-constraining. Nevertheless, for smaller $\mchi$ the resulting lower
bound on $\maxino$ will indeed be typically much stronger than the
above bounds.  For example, for a bino of $60\gev$ one finds
$\maxino\gsim 360\mev$. However, increasing $\mchi$ to $150\gev$
removes the bound altogether.

We should mention other non-thermal mechanisms for re-populating the
Universe with axinos and/or other relics such as gravitinos and
moduli.  First, it has recently been claimed~\cite{grt} that
gravitationally interacting particles can be copiously produced
through inflaton field oscillations in the reheating or preheating
process, and that non-thermal production of gravitinos can be much
more efficient than through thermal processes.\footnote{This claim has
  very recently been disputed~\cite{nps01}. See also
  Ref.~\cite{grt01}.}  This would further aggravate the gravitino
problems with nucleosynthesis, thus leading to a much more severe
constraint on the reheating temperature, sometimes as low as
$\treh\lsim 10^5\gev$, depending on the model. We do not expect the
mechanism to contribute to the production of axinos in any significant
way, especially in the regime of low $\treh$ mentioned above.
Furthermore, such mechanism would strongly depend on the inflationary
model and on the interactions between the inflaton, and the axino and
it would seem difficult, if not impossible, to obtain the required
abundance of DM.

Second, axinos can be produced in gravitino decays, along with their
non-SUSY partner, the axion~\cite{ay00}.  It is worth emphasizing that
the gravitino problem can be easily resolved if the axino is the LSP
and the hierarchy of masses $\mlosp>\mgravitino>\maxino$ is assumed.
The process will take place long after nucleosynthesis ($t\sim10^8\,
{\rm sec}$) but decay products will now be completely harmless. The
lightest ordinary superpartner will also decay directly to axinos,
thus by-passing the dangerous late decays to gravitinos and energetic
photons.  In the relatively low $\treh\lsim10^5\gev$ regime, where NTP
of axinos dominates, there is not even the gravitino problem
anymore~\cite{grt}.  However, the mechanism resolves the gravitino
problem even at much larger $\treh\lsim10^{15}\gev$ studied in
Ref.~\cite{ay00}. In other words, it seems that the gravitino problem
can be resolved, assuming the above hierarchy, in a model-independent
way. At large $\treh$ the axino would not contribute to cold DM.
Instead, it would be warm DM while cold DM could be provided by, for
instance, the axion.

It is clear that in discussing cosmological implications of axinos one
quantity of crucial importance is the axino mass itself.
Unfortunately, unlike the axino coupling, this quantity is rather
poorly determined and is strongly model-dependent. As mentioned above,
a distinctive feature of axinos is that typically their mass is not set by the
SUSY-breaking scale and therefore can be much lower. In fact, it can
span a wide range from $\sim\ev$ to $\sim\gev$~\cite{ckn}, depending
on the model.  In this study we will therefore treat $\maxino$ as a
basically free parameter.

Finally, we briefly mention the saxion -- the $R=1$ spin-zero
component of the axion supermultiplet.\footnote{ In the literature $s$
is called saxion ($\equiv$ scalar partner of axion)~\cite{rtw}, or
saxino ($\equiv$ the scalar partner of axino)~\cite{kim91}.  Since
superpartner names ending with ``-ino'' seem to be reserved for
(Majorana) fermions, here we will use the name `saxion'.} As with other
scalars, the mass of the saxion arises from a soft term and is
typically of the order of the SUSY-breaking scale. Cosmological
properties of saxions can be quite important~\cite{ck,kim91}. In
particular, saxions decay relatively fast which may cause significant
entropy production which would reheat the Universe and lead to other
consequences. However, these effects will not be relevant to us if the
saxion mass is below $1\tev$ which we will assume here. We
will occasionally comment on the saxion below when relevant for axino
cosmology.

The plan of the paper is as follows. In Section~\ref{sec:axino} we
discuss relevant axion and axino properties. In particular, we
concentrate on the axino mass and couplings. In Section~\ref{sec:TP}
the thermal production of axinos is analyzed in detail.
Section~\ref{sec:NTP} summarizes and extends the previously obtained
results in the case of NTP~\cite{ckr}. Astrophysical and cosmological
constraints are discussed in Section~\ref{constraints:sec}. In
Section~\ref{sec:TPvsNTP} the two production mechanisms are compared
and further discussed. Finally, in Section~\ref{sec:implications},
implications for cosmology and for collider phenomenology are briefly
discussed and concluding remarks are made.

%%%%%%%%%%%%%%%%%%%%%%%%%%%%%%%%%%%%%%%%%%%%%%%%%%%%%%%%%%%%%%%%%%%%%%%%%%%
\section{The axino}
\label{sec:axino}

First, let us briefly summarize the main properties of axions.
The axion can be defined as a pseudo-scalar
particle with the effective interaction given by
\begin{equation}
{\cal L}_{a}^{\rm eff} = \frac{a}{M} F \widetilde F,\label{adefinition}
\end{equation}
where $M$ is a model-dependent mass scale, $F$ is the field strength of the
gluon field and $\widetilde F$ is its dual. The potential arising from
the above interaction settles $\langle a\rangle$ to zero.  In general,
there can arise some other small terms in the axion potential in
addition to the one arising from (\ref{adefinition}).  (When CP
violation is considered $\langle a\rangle=0$ is shifted, but the shift
is extremely small since a square of the weak interaction coupling and
the pion mass appear in the resulting expression~\cite{gr}.)  At low
energy, one may not question how $M$ is generated. It can arise in
renormalizable or non-renormalizable theories, or in composite
models. If a fundamental theory such as superstrings allows the above
non-renormalizable interaction, then $M$ is the compactification
scale~\cite{ssaxion}. If a confining force at a scale $\Lambda_a$
produces the Goldstone boson with the above coupling at low energy,
then $M$ is of order $\Lambda_a$ and the axion is
composite~\cite{compaxion}. If the axion resides in the complex Higgs
multiplet(s), it can be derived through the PQ mechanism with the
spontaneously broken chiral $U(1)$ global symmetry.

In phenomenologically acceptable axion models, the mass scale $M$ 
is given by $M = 8 \pi f_a/\alpha_s $, where $f_a$ is the PQ scale, 
$10^9\gev\lsim f_a\lsim 10^{12}\gev$. In
these models the axion mass is given by
$m_a\sim\Lambda_{QCD}^2/f_a\sim10^{-2}$--$10^{-5}\ev$; in other words 
the axion is very light.
These models can
be classified as hadronic axion models (usually called the KSVZ
models~\cite{ksvz}) and DFSZ axion models~\cite{dfsz}. Let us briefly
recall some basic features of these two most popular implementations
of the PQ mechanism. (Detailed presentations can be found in several
excellent reviews, \eg\ in~\cite{axionreviews:cite}.) In both one
assumes the existence of a complex scalar field $\phi$, which is a
singlet under the SM gauge group but carries a PQ charge.  When $\phi$
develops a VEV, $\langle \phi \rangle = \fa$, the PQ symmetry gets
broken and the phase of $\phi$ becomes the axion field. To relate the
global charge of the complex scalar $\phi$ to the PQ charge (i.e. to
couple the axion to the gluon anomaly), in the KSVZ scheme~\cite{ksvz}
one further introduces at least one heavy quark $Q$~\footnote{
In general one can introduce, instead of a single colored fermion,
a multiplet, \eg\  a non-singlet vector-like quark representation
of $SU(2)\times U(1)_Y$.}
which couples to $\phi$ through the term
\begin{equation}
{\cal L}_{PQ} = f_Q \phi {\bar Q}_R Q_L + {\rm h.c.}
\label{lksvz:eq}
\end{equation}
where $f_Q$ is the Yukawa coupling. 
The PQ charge of $\phi$ is $+2$, while
those of $Q_{R,L}$ are $\pm1$. 
The breaking of the PQ symmetry gives large mass to the heavy quark,
$m_Q=f_Q \fa$. The axion interacts with ordinary matter through loops
involving the exchange of the heavy quark (i.e. through the anomaly term). 

This leads, below the PQ scale, to effective axion interactions 
terms; among these the most
important one is its interaction with gluons
\begin{equation}
{\cal L}_{agg} = {\alpha_s \over 8\pi\fa} a F \widetilde F.
\label{effcoup:eq}
\end{equation}

In the DFSZ model~\cite{dfsz}, instead of the heavy quark, one
introduces two Higgs doublets~\cite{pq,axion} $\phi_{u,d}$ of
$SU(2)_L\times U(1)_Y$ which couple the SM quark sector to $\phi$.
The doublets carry hypercharges $\pm 1/2$ and PQ charges $-Q_{u,d}$,
respectively. The $\phi_u$ field couples only to up-type quarks, while
$\phi_d$ couples only to down-type quarks and leptons which carry PQ
charges so as to respect the PQ symmetry.

After the breaking of the PQ symmetry, again effective axion
interactions with ordinary matter are generated, including the
term~(\ref{effcoup:eq}).  In both models there appears an additional
quantity, the number of different vacua (or, equivalently, domain walls)
$N$: $N=1(6)$ for the KSVZ
(DFSZ) model. For our purpose it is enough to note that its effect
will be to replace $\fa$ by $\fa/N$.

Both models can be readily
supersymmetrized~\cite{susyaxion1,susyaxion2}.  The
scalar field $\phi$ becomes promoted to the superfield, and
accordingly the other fields as well.  In particular, the axion, being
the phase of $\phi$, is also promoted to an axion supermultiplet
$\Phi$ consisting of the pseudo-scalar axion $a$, its fermionic partner,
the axino $\axino$, and the scalar partner, the saxion $s$
\begin{equation}
\label{Phi:eq}
\Phi={1\over\sqrt{2}}\left(s+ia\right)
+\sqrt{2}\tilde a\theta+F_\Phi\theta\theta.
\end{equation}
The axino is thus a neutral, $R=-1$, Majorana chiral fermion. 
Adding supersymmetry opens up a wide range of choices
for the implementation and breaking of the PQ symmetry
through a choice of different superpotentials and SUSY-breaking schemes,
as will be illustrated below. 

Now let us move on to the discussion of axino masses.
We will briefly summarize the relevant results of several previous
studies~\cite{tw,kmn,rtw,Goto-Yamaguchi,ckn,cl}. The overall
conclusion will be that the mass of the axino is strongly
model-dependent~\cite{rtw,ckn}. It can be very small ($\sim\ev$), or large
($\sim\gev$), depending on the model.

It is worth emphasizing that, unlike the case of the gravitino and
ordinary superpartners, the axino mass does not have to be of the order of
the SUSY-breaking scale in the visible sector,
$\msusy$~\cite{tw,rtw,ckn}. In global SUSY models, one
sets $\msusy={\cal O}(1\tev)$ on the basis of naturalness. In SUGRA
models $\msusy$ is set by the gravitino mass
$\mgravitino\sim{\ms^2/\mplanck}$, where $\ms\sim10^{11}\gev$ is the
scale of local SUSY-breaking in the hidden sector and $\mplanck \simeq
2.4\times10^{18}\gev$ denotes the Planck mass.

It is easy to see why $\maxino$ is not generically of the order of
$\msusy$. In the case of unbroken SUSY, all members of the axion
supermultiplet remain degenerate and equal to the tiny mass of the
axion given by the QCD anomaly.  Once SUSY is broken, the saxion,
being a scalar, receives a soft-mass term $\msaxion^2
s^2$~\cite{tw,nieves}, where $\msaxion\sim \msusy$, similarly to the
other scalar superpartners. 

Of course, the axion, being instead a phase of the fields whose VEVs
break the PQ symmetry, does not receive soft mass.  Likewise, one
cannot write a {\em soft} mass term for the axino 
since it is a chiral fermion
(for the same reason there are no soft terms for, for
instance, the MSSM higgsinos) and a superpotential term 
$W\sim{\rm(mass\ parameter)}\cdot \Phi\Phi$ is  absent due 
to the PQ symmetry. The lowest-order term one can write
will be a non-renormalizable term of dimension-5. The axino mass
will then be of order $\msusy^2/\fa\sim1\kev$~\cite{tw}. (A more
detailed, and more sophisticated, explanation can be found, \eg, in
Ref.~\cite{ckn}.)

However, in addition to this inevitable source of axino mass, one can
easily generate much larger contributions to $\maxino$ at one-loop or
even at tree-level.  One-loop terms will always contribute but will
typically be $\lsim\msusy$ (KSVZ) or even $\ll\msusy$ (DFSZ).
Furthermore, in non-minimal models where the axino mass eigenstate
comes from more than one superfield, $\maxino$ arises even at the
tree-level. In this case $\maxino$ can be of order $\msusy$ but can
also be much smaller.  These points are illustrated by the following
examples.

First let us see how tree-level contributions can be generated from
superpotentials involving singlet fields.  For example, consider the
case where the PQ
symmetry is assumed to be broken by the renormalizable
superpotential term (in KSVZ or DFSZ models)~\cite{jekim83}
\begin{equation}
\label{kim83} 
W=fZ(S_1S_2-\fa^2),
\end{equation}
where $f$ is a coupling, and $Z, S_1$ and $S_2$
are chiral superfields with PQ charges of $0$, $+1$ and $-1$, 
respectively. In this case the axino mass can be at the soft SUSY-breaking 
mass scale~\cite{Goto-Yamaguchi}. It  arises from diagonalizing the mass
matrix of the fermionic partners $\tilde S_1,\tilde S_2$, and $\tilde Z$,
\begin{equation}
\label{axmass}  
\left(\begin{array}{ccc}
0 & m_{\axino} & f\fa \\
m_{\axino} & 0 & f\fa \\
f\fa & f\fa & 0
\end{array}\right),
\end{equation}
where $m_{\axino}=f\langle Z\rangle$
and $ff_a\sim10^{11}\gev$.
The mass matrix in Eq.~(\ref{axmass}) is
for three two-component neutral fermions.
These three will split into one Dirac fermion, and one Majorana fermion
which is interpreted as the axino.
The corresponding eigenvalues are $\lambda=-\maxino$ and
$\lambda=\pm\sqrt{2}ff_a+{\cal O}(\maxino)$.  In the global SUSY
limit, $\langle Z\rangle=0$ and the tree-level axino mass would be
zero.  However, when $S_1$ and $S_2$ acquire VEVs and soft terms are
included, $V=|f|^2(|S_1|^2+|S_2|^2)|Z|^2+(A_1 f S_1 S_2 Z- A_2 f f_a^2
Z+ {\rm h.c.})$, a linear term in $Z$ is generated which induces
$\langle Z\rangle$ of order $(A_1-A_2)/f$ where $A_{1,2}$ are the soft
trilinear mass parameters. The axino mass thus arises
at the soft mass scale~\cite{Goto-Yamaguchi}.

However, by choosing a more complicated superpotential, one can
significantly lower the axino mass~\cite{ckn} even at the tree-level.
Consider a supergravity superpotential consistent with the PQ symmetry
as
\begin{equation}
W^\prime=fZ(S_1S_2-X^2)+\frac{1}{3}\lambda(X-M)^3,
\label{ckn}
\end{equation}
where $X$ carries a zero PQ charge. In this case a minimization of the
potential resulting from $W^\prime$ is much more complicated, and an
approximate solution gives the lightest eigenvalue of the fermion mass
matrix with $\langle V\rangle=0$ which is $m_{\tilde a}={\cal
  O}(A-2B+C)+{\cal O}(m_{3/2}^2/f_a)$, where $A$, $B$ and $C$ are the
respective trilinear, bilinear and linear soft breaking parameters.
For the standard pattern of soft breaking terms, $B=A-m_{3/2}$ and
$C=A-2m_{3/2}$, the leading contribution $A-2B+C$ vanishes and 
the tree-level axino mass becomes of order $m_{3/2}^2/f_a\sim1\kev$.

The above example shows that in calculating even the tree-level
axino mass one must carefully consider the full potential $V$
generated in supergravity models. In general, the tree-level axino
mass either can be of order $m_{3/2}$ or, depending on the PQ sector
of the model and the pattern of soft breaking parameters, can  be
much less, as shown in the above examples.  The detailed conditions
for this to happen were analyzed by Chun and Lukas~\cite{cl}.
In particular one can also have a more complicated PQ sector and 
break the PQ symmetry by non-renormalizable terms in the superpotential 
leading to interesting cosmological implications~\cite{ckl00}. 

If the tree-level axino mass is either zero or of order
$m_{3/2}^2/f_a$, the contribution from loop diagrams can become more
important.  In the global SUSY version of the KSVZ model, axino mass
will arise at the one-loop level with a SUSY-breaking $A$-term
insertion at the intermediate heavy squark line. Then one finds
$\maxino \sim (f_Q^2/8\pi^2)A$, where $f_Q$ is the Yukawa coupling of
the heavy quark to a singlet field containing the axion (compare
Eq.~(\ref{lksvz:eq})), which gives $\maxino$ in the range of $\lsim$ a few tens
of GeV~\cite{Moxhay-Yamamoto,Goto-Yamaguchi}.  In the DFSZ
model~\cite{dfsz} (in either global SUSY or supergravity), where such
$A$-term contribution is absent, the axino mass remains in the $\kev$
range, as was pointed out in~\cite{kmn,rtw}. Thus, in some models
the axino mass can be rather small.

In gauge-mediated SUSY-breaking models (GMSB), the pattern of the
axino mass is completely different.  In the GMSB approach, the
tree-level effect of supergravity is not important since the gravitino
mass is much smaller than soft breaking masses, typically of order
$\sim\ev$ to $\sim\kev$. These models attracted much interest in the
recent past since the gravitino production rate is quite large which
leads to important implications for cosmology~\cite{kk} and
accelerator experiments~\cite{large}. In the GMSB models, the axino mass
is further reduced by one more power of $f_a$ and can range from
$10^{-9}$~eV to 1~keV~\cite{chun}. 
It remains model-dependent but again can be smaller than 
the gravitino mass. We will not discuss GMSB cases here any more.

In conclusion, we see that a complete knowledge of the superpotential and
supersymmetry-breaking mechanism is necessary to pin down the axino mass. 
In general it can range from $\sim\ev$ to $\sim\ {\rm tens\ of}\ \gev$. 
For the sake of
generality, in discussing cosmological properties of axinos, we will
use $\maxino$ as a free parameter.

In addition to its mass, the axino couplings to gauge and matter
fields are of crucial importance for the study of its cosmological
abundance. In general, the couplings of the members of the axion/axino
supermultiplet will be determined by the PQ symmetry and
supersymmetry. The most important term for our purpose will be the
coupling to gauge multiplets.  It is given by the same interaction term
as the one which gives rise to the ``$\theta$'' term in the QCD Lagrangian. In
supersymmetric notation it can be written as
\begin{equation}
\label{axionints:eq}
{\cal L}= {\alpha_Y\cayy \over 4\sqrt{2}\pi\left(\fa/N\right)}
(\Phi B_\alpha B^\alpha)_{\theta\theta} 
+ {\alpha_2\caww \over 4\sqrt{2}\pi\left(\fa/N\right)} [B \ra W]
+ {\alpha_s \cagg \over 4\sqrt{2}\pi\left(\fa/N\right)} [B \ra F]  +
{\rm h.c.},
\end{equation}
where $\alpha_i$ ($i=Y,2,s$) are the coupling constant strengths of
the SM subgroups, and $B$, $W_3$ and $F$ are the respective gauge
supermultiplets.  The coefficient $\cagg=1$ is universal, while $\cayy$
and $\caww$ are model-dependent.  For processes that involve the
electroweak gauge bosons and superpartners, it is always possible to
rotate away the interaction of the axion with the $SU(2)_L$ gauge
multiplet through some anomalous global chiral $U(1)_X$ 
rotation  since the PQ symmetry is global.
Then the only interaction that remains to be specified will be the one 
between the axino and the $U(1)_Y$ gauge multiplet.
This will be equivalent to assigning PQ charges to
left-handed doublets in such a way as to obtain $C_{aWW}=0$. One will then
also find $C_{aYY}=C_{a\gamma\gamma}$.  

The rationale for removing the coupling $C_{aWW}$ is as follows.  We
would like to keep the gauge couplings and top quark Yukawa coupling
but neglect the other Yukawa couplings. This is because the neglected
Yukawa couplings will not be important in the production cross
section. Moreover we expect that the lightest neutralino is likely to
contain a significant bino component~\cite{chiasdm,rr93,kkrw}, so that
phenomenologically the interaction between the $U(1)_Y $ field
strength and the axion multiplet will be important. Keeping this in mind,
we will work with the simplest interactions and use the basis where
the $C_{aWW}$ coupling is absent.

The procedure of global $U(1)_X$ rotation and charge redefinition is
somewhat subtle and requires a word of clarification on which we
digress here. This rotation applies only to the high energies of interest
to us, but not to the low-energy phenomena below the QCD scale.
We can use a global $U(1)_X $ rotation to redefine the 
PQ charges of the quarks like $Q^\prime=Q- Q({\rm quark\ doublet})$.
This will cancel the PQ charge of the quark doublets, 
but will also shift the PQ charges of the other fields.
Since the $Q^\prime$ charges of the quark doublets are
absent, one would expect that 
there is no $C_{aWW}$ coupling either. However, there will be 
(in DFSZ models) lepton and Higgs doublets whose 
$Q^\prime$ charges will in general be non-zero. Thus the $Q^\prime$
charge should instead be defined by the rotation removing $C_{aWW}$,
in which case $Q^\prime$ charges of quark doublets and lepton 
and Higgs doublets may in general be different from zero. 
Because the top quark Yukawa coupling is kept, the global
$U(1)_X$ transformation must either leave the quark fields
unchanged or operate on both the left-handed and 
right-handed quarks and the Higgs field, similarly to the 
Standard Model $U(1)_Y$. 

The rotation will not change the $\cagg$ coupling 
since $U(1)_Y $ is anomaly-free. 
On the other hand, for leptons the Yukawa coupling can be neglected 
and we can therefore use any global $U(1)_X$ 
such that the final $C_{aWW}$ coupling is absent.
The above rotation amounts to changing the PQ charges of the
lepton doublets and hence to modifying all the couplings 
of the $a$-field with the field strengths of $W_\mu$ and  $B_\mu$.
If we could neglect also the top quark Yukawa coupling, 
we could consider instead a basis without the $\cagg$ 
coupling where the interaction~(\ref{adefinition}) would be absent and
no axion field could be defined.
But that scenario would correspond to the decoupling of the axion
from the SM sector which takes place in the massless quark case.

In conclusion, the axion coupling must be defined rigorously at the
axion vacuum. However, above the TeV scale we can choose any basis for 
calculating the axino production if the contribution for $E\lsim1\tev$  is not
important. 

After the $U(1)_X$ rotation, the $SU(2)$ singlets (quarks and leptons) 
carry $Q^\prime$ charges
and hence the $aYY$ coupling will not be absent:
\begin{equation}
{{\alpha_Y \cayy}\over{\fa/N} } a B\widetilde B,
\end{equation}
where $ B$ denotes here the field strength of the $U(1)_Y$ gauge boson
$B_\mu$ and $\widetilde B$ denotes its dual.
Since $B_\mu= A_\mu \cos\theta_W + Z_\mu \sin\theta_W$, we obtain
\begin{equation}
{{\alpha_Y \cos^2\theta_W\cayy }\over{\fa/N} } a F_{em}\widetilde F_{em},
\end{equation} 
where, $ F_{em}$ is the field strength of the electromagnetic gauge boson
and $\widetilde F_{em}$ is its dual. This implies $\cayy=C_{a\gamma\gamma}$.

In the DFSZ model with $(d^c,e)$-unification,
$C_{aYY}=8/3$. In the KSVZ model for $e_Q=0,-1/3,$ and 2/3,
$C_{aYY}=0,2/3$ and 8/3, respectively~\cite{coupling}. Below the QCD
chiral symmetry-breaking scale, $C_{a\gamma\gamma}$ and $C_{aYY}$ are
reduced by $1.92$.  

For our purpose the most important coupling will be
that of axino--gaugino--gauge boson interactions
which can be derived from Eq.~(\ref{axionints:eq}) and written in a more
conventional way as a dimension-5 term in the Lagrangian:
\begin{equation}
{\cal L}_{\axino\lambda A} = i \frac{\alpha_Y \cayy}{16\pi\left(\fa/N\right)} 
{\bar{\axino}} \gamma_5 [\gamma^\mu,\gamma^\nu]\bino B_{\mu\nu}
+ i \frac{\alpha_s}{16\pi\left(\fa/N\right)}
{\bar{\axino}}\gamma_5[\gamma^\mu,\gamma^\nu]\gluino^b F^b_{\mu\nu}.
\label{eq:axino-gaugino-gauge}
\end{equation}
Here and below $\bino$ denotes the bino, 
the fermionic partner of the $U(1)_Y$
gauge boson $B$ and $\gluino$ stands for the gluino.

One can also think of terms involving dimension-4 operators
coming, \eg, from the {\em effective} superpotential $\Phi\Psi\Psi$
where $\Psi$ is one of the MSSM matter (super)fields.
However, axino production processes coming from such terms will be
suppressed at high energies with respect to processes involving
Eq.~(\ref{eq:axino-gaugino-gauge}) by a factor $m_\Psi^2/s$, where $s$
is the square of the center of mass energy.  We will comment on the
role of dimension-4 operators again below but, for the most part,
mostly concentrate on the processes involving axino interactions with
gauginos and gauge bosons, Eq.~(\ref{eq:axino-gaugino-gauge}), which are
both model-independent and dominant.

%%%%%%%%%%%%%%%%%%%%%%%%%%%%%%%%%%%%%%%%%%%%%%%%%%%%%%%%%%%%%%%%%%%%%%%%%%%
\section{Thermal production}
\label{sec:TP}

As stated in the Introduction, thermal production proceeds through
collisions and decays of particles that still remain in the
thermal-bath. As we will see, its efficiency will strongly depend on
effective interaction strengths of the processes involved and on
characteristic temperatures of the bath.  Particles like axinos and
gravitinos are somewhat special, in the sense that their interactions
with other particles are very strongly suppressed with respect to the
Standard Model interaction strengths.  Therefore such particles remain
in thermal equilibrium only at very high temperatures. In the
particular case of axinos (as well as axions and saxions), 
their initial thermal
populations decouple at~\cite{rtw}
\begin{equation}
\label{eq:T-axino-decouple}
\Td\sim 10^{10}\gev\,\left(\frac{\fa/N}{10^{11}\gev}\right)
\,\left(\frac{\alphas}{0.1}\right)^{-3}.
\end{equation}
At such high temperatures, the axino number density is the
same as the one of photons and other relativistic species.  In other
words, such primordial axinos freeze out as {\it relativistic}
particles. Rajagopal, Turner and Wilczek (RTW)~\cite{rtw} pointed out
that, in the absence of a subsequent period of inflation, the
requirement that the axino energy density be not too large
($\omegaa\lsim1$) leads to

\beq 
\maxino <
12.8\ev\left({\gstar(\Td)\over\geff}\right),
\label{eq:rtw:warmaxinobound}
\eeq 
where $\geff=1.5$ and $\gstar$ is the number of effectively
massless degrees of freedom (particles with mass much smaller than
temperature). In the MSSM at  temperatures much
higher than $\msusy$, one has
$\gstar=915/4$. The RTW bound~(\ref{eq:rtw:warmaxinobound}) then takes a
more well-known form~\cite{rtw} 
\beq 
\maxino < 2\kev,
\label{eq:rtw:warmaxino}
\eeq 
and the corresponding axinos would be light and would provide
warm or even 
hot dark matter~\cite{rtw}. We will not consider this case in the
following, primarily because we are interested in cold DM axinos.  We
will therefore assume that the initial population of axinos (and other
relics, such as
gravitinos), which were 
present in the early Universe, was subsequently diluted
away by an intervening inflationary stage and that the reheating
temperature after inflation was smaller than $\Td$. It also had to be
less than $\fa$, otherwise the PQ symmetry would have been restored, thus
leading to the well-known domain-wall problem associated with global
symmetries. 

At lower temperatures, $T<\Td$, the Universe can be re-populated with
axinos (and gravitinos) through scattering and decay processes
involving superpartners in the thermal-bath.  As long as the axino
number density $\naxino$ is much smaller than $\ngamma$, the
number density of photons in thermal equilibrium, its time evolution
will be adequately described by the Boltzmann equation:
\begin{equation}
\label{eq:Boltzmann}
\frac{d\,\naxino}{d\,t} + 3H \naxino =
\sum_{i,j} \langle\sigma(i+j\ra \axino+\cdots)\vrel\rangle n_i n_j +
\sum_i \langle\Gamma(i\ra \axino+\cdots)\rangle n_i.
\end{equation}
Here $H$ is the Hubble parameter,
$H(T)=\sqrt{\left(\pi^2\gstar\right)/\left(90\mplanck^2\right)}\;T^2$, 
where $g_*$ has been defined above and 
$\sigma(i+j\ra \axino+\cdots)$ is the scattering cross section for
particles $i,j$ into final states involving axinos,
$\vrel$ is their relative velocity,
$n_i$ is the $i$th particle number density in the thermal-bath,
$\Gamma(i\ra \axino+\cdots)$ is the decay width of the $i$th particle,
$\langle\cdots\rangle$ stands for thermal averaging. (Averaging over initial
spins and summing over final spins is understood.)
Note that on the {\it r.h.s.} we have neglected inverse processes,
since they are suppressed by $\naxino$.

\underline{Solving the Boltzmann Equation.} \hspace{0.2cm}
In order to solve the Boltzmann Eq.~(\ref{eq:Boltzmann}),
it is convenient to introduce the axino TP yield
\begin{equation}
\ythaxino = \frac{\naxino^{\rm TP}}{s},
\label{ythaxino:eq}
\end{equation}
where 
$s= (2\pi^2/45)\gsstar T^3$ is the entropy density, 
and normally $\gsstar=\gstar$ in the early Universe. 
We also change variables from the cosmic time $t$ to the temperature $T$ 
by $-dt/dT=1/HT$ for the radiation dominated era.
One integrates the Boltzmann equation from the reheating temperature
$\treh$ after inflation down to zero.  The yield can be written as
\begin{equation}
\ythaxino = \sum_{i,j}\yascat + \sum_{i}\yadec,
\end{equation}
where the summation over indices $i,j$ is the same as in
Eq.~(\ref{eq:Boltzmann}).
The expressions for $\yascat$ and $\yadec$ are given by
\begin{equation}
\label{eq:Yscat}
\yascat =
\int_{0}^{\treh}dT\,
\frac{\langle\sigma(i+j\rightarrow\axino +\cdots)\rangle n_in_j}{sHT}
\end{equation}
and
\begin{equation}
\label{eq:Ydec}
\yadec = 
\int_{0}^{\treh}dT\,
\frac{\langle\Gamma(i\rightarrow\axino +\cdots)\rangle n_i}{sHT}.
\end{equation}
Explicit formulae for $\yascat$ and $\yadec$ can be found, for instance, in
Ref.~\cite{chkl}.

\underline{Scatterings.} \hspace{0.2cm} In the case of axinos the main
productions channels are the scatterings of (s)particles described by
a dimension-5 axino--gaugino--gauge boson term in the
Lagrangian~(\ref{eq:axino-gaugino-gauge}).  Because of the relative
strength of $\alphas$, the most important contributions will come from
strongly interacting processes.  We will discuss them first.

The relevant 2-body processes for
strongly interacting particles $i,j$ into several final states involving
axinos are listed in Table~1. %\ref{table:cross-section}. 
The corresponding 
cross sections $\sigma_n=\sigma (i+j\ra \axino+\cdots)$, where the
label $n=A,\ldots,J$ counts the different allowed combinations of the
initial and final-state particles, can be written as
\begin{equation}
\label{eq:cross-section}
\sigma_n(s) = \frac{\alphas^3}{4\pi^2\left(\fa/N\right)^2}{\sigmabar}_n(s)
\end{equation}
where $\sqrt{s}$ is the energy in the center-of-mass frame. 
Also listed there are the respective: spin factor $\nspin$ (the
number of spin combinations in the initial state), flavor factor
$\nflavor$ (number of color-triplet chiral multiplets), and number
density factor $\eta_i$, $i=1,2$
($\eta_i=3/4\; (1)$ for each initial-state fermion (boson).
The group factors $f^{abc}$ and $T^{a}_{jk}$ of the gauge group
$SU(N)$ satisfy the usual relations $\sum_{a,b,c}|f^{abc}|^2=N(N^2-1)$ and
$\sum_{a}\sum_{jk}|T^{a}_{jk}|^2=(N^2-1)/2$.

\begin{table}
\label{table:cross-section}

\begin{center}
\begin{tabular}{|c||c|l|c|c|c|} \hline
n & Process & \makebox[50mm][c]{$\sigmabar_N$} \hfill & $\nspin$
& $\nflavor$ & $\eta_1\eta_2$
\\\hline\hline
A & $ g^a + g^b \ra \tilde a + \tilde g^c $ &
$\frac{1}{8}|f^{abc}|^2$ & 4 & 1 & 1
\\\hline
B & $ g^a + \tilde g^b \ra \tilde a + g^c $ & 
$\frac{5}{16}|f^{abc}|^2\left[\logsoms-\frac{15}{8}\right]$ & 4 & 1 &
$\frac{3}{4}$
\\ \hline
C & $ g^a + \tilde q_k  \ra \tilde a + q_j $ &
$\frac{1}{8}|T^{a}_{jk}|^2$ & 2 & $N_F\times 2$ & 1
\\ \hline
D & $ g^a + q_k  \ra \tilde a + \tilde q_j $ &
$\frac{1}{32}|T^{a}_{jk}|^2$ & 4 & $N_F\times 2$ &$\frac{3}{4}$
\\ \hline
E & $ \tilde q_j + q_k  \ra \tilde a + g^a $ &
$\frac{1}{16}|T^{a}_{jk}|^2$ & 2 & $N_F\times 2$ &$\frac{3}{4}$
\\ \hline
F & $ \tilde g^a + \tilde g^b \ra \tilde a + \tilde g^c $ &
$\frac{1}{2}|f^{abc}|^2\left[\logsoms-\frac{29}{12}\right] $ & 4 & 1 &
$\frac{3}{4}$  $\frac{3}{4}$
\\ \hline
G & $ \tilde g^a + q_k \ra \tilde a + q_j $ &
$\frac{1}{4}|T^{a}_{jk}|^2\left[\logsoms-2\right]$ & 4 & $N_F$ &
$\frac{3}{4}$  $\frac{3}{4}$
\\ \hline
H & $ \tilde g^a + \tilde q_k \ra \tilde a + \tilde q_j $ & 
$\frac{1}{4}|T^{a}_{jk}|^2\left[\logsoms-\frac{15}{8}\right]$ & 2 &
$N_F\times 2$ &$\frac{3}{4}$
\\ \hline
I & $ q_k + {\bar q_j} \ra \tilde a + \tilde g^a $ &
$\frac{1}{24}|T^{a}_{jk}|^2$ & 4 & $N_F$ &
$\frac{3}{4}$  $\frac{3}{4}$
\\ \hline
J & $ \tilde q_k + \tilde q_j \ra \tilde a + \tilde g^a $ &
$\frac{1}{24}|T^{a}_{jk}|^2$ & 1 &
$N_F\times 2$ & 1
\\ \hline
\end{tabular}
\end{center}

%%% Local Variables: 
%%% mode: latex
%%% TeX-master: t
%%% End: 

\caption{The cross sections for the different axino thermal-production 
channels involving strong interactions. Masses of particles
have been neglected, except for the plasmon mass $\meff$. 
See text for explanation of the different symbols.}
\end{table}

\EPSFIGURE{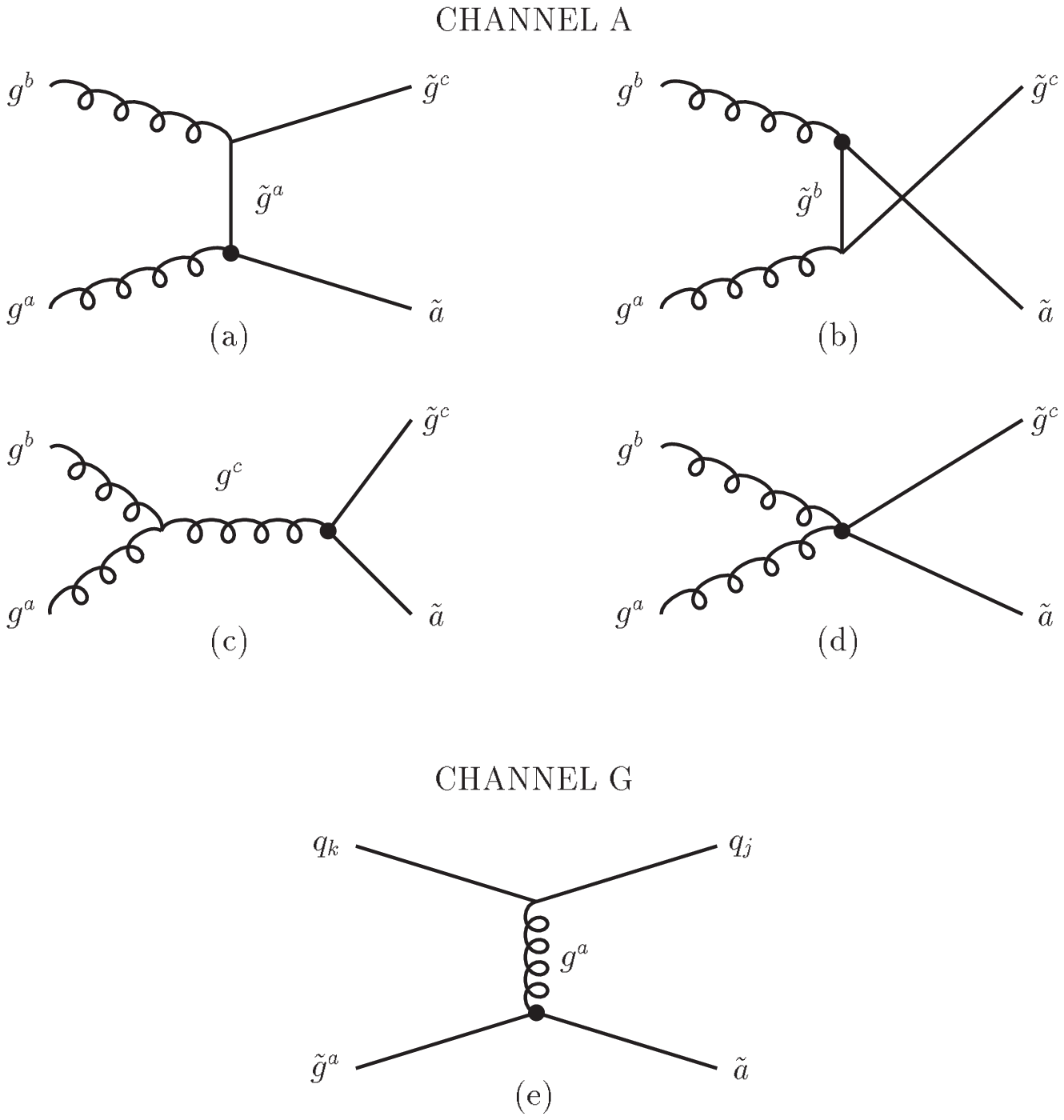,width= 5.25in}
{Feynman diagrams contributing to channels A and G in
    Table~1. Thick dot denotes effective axino 
couplings to ordinary colored
    particles.  
\label{fig:ckkr-fdiag1} 
}

The diagrams listed in the Table are analogous to those involving
gravitino production considered by Moroi \etal~\cite{mmy}, and later
by Bolz \etal~\cite{bbp98}, and we follow their classification.  This
analogy should not be surprising, since both particles are neutral
Majorana superpartners. (We will return to the gravitino--axino
analogy in Section~\ref{sec:TPvsNTP}.) In Fig.~\ref{fig:ckkr-fdiag1}
we present the Feynman diagrams corresponding to channels~A and G.
The diagrams for the other channels can be drawn in a similar fashion.

The diagrams in channels B, F, G and H involve $t$-channel
exchange of massless gluons and are therefore divergent. In order to
provide a mass regulator, one can introduce a plasmon mass $\meff$
representing the effective gluon mass due to plasma effect. We follow
the prescription of Ref.~\cite{ekn84}, which has also
been used in Refs.~\cite{mmy,bbp98}, and assume $\meff^2= g_s^2 T^2$
where $g_s$ is the strong coupling constant.

Note that we have neglected in the gluon thermal mass a potentially
sizeable factor which counts the number of colored degrees of
freedom of the plasma but kept a constant term in the
logarithmically divergent cross sections, which depends on the
procedure used to regulate the divergence.  While a more proper
treatment exists in the literature~\cite{by91,wil}, based on Thermal Field
Theory and taking into account also the Fermi or Bose nature of the
particles in the thermal-bath, we believe that the above approximation
is good enough for our purposes. This belief is supported in
particular by the fact that in the case of gravitinos, the full
treatment agrees with an approximate one within a factor 3
\cite{wil}. For simplicity, in computing $\sigmabar_n(s)$ we
neglected all mass terms (involving gauginos, axinos and scalars) and
kept only the plasmon mass $\meff^2$.

Finally, we comment on the scattering processes involving the
hypercharge multiplet. It is clear that these will always be
subdominant. This is not only a result of a weaker interaction
strength relative to gluons and gluinos (so long as $C_{aYY}$ is not
too large), but is also caused by a smaller number of production channels.
Indeed, since $U(1)_Y$ is Abelian, in this case many partial cross
sections (all channel A, B and F contributions)
vanish automatically.  The others are given by the same
expressions as in Table~1, with the substitution
$\alpha^3_s\rightarrow\alpha^3_YC^2_{aYY}$ and
$|T^a_{ij}|^2\rightarrow Y^2_i$,
where $Y_i$ is the hypercharge of the initial (and final) state.

\underline{Decays.} \hspace{0.2cm} In addition to scattering
processes, axinos can also be produced through decays of heavier
superpartners in thermal plasma. At temperatures $T\gsim\mgluino$,
these are dominated by the decays of gluinos into LSP axinos and
gluons.  The relevant decay width is given by
\begin{equation}
\label{eq:Gamma-gluino}
\Gamma(\gluino^a\ra \axino + g^b) = \delta^{ab}
\frac{\alpha_s^2}{128\pi^3}\frac{\mgluino^3}{\left(\fa/N\right)^2}\left(1-
\frac{\maxino^2}{\mgluino^2}\right)^3
\end{equation}
and one should sum over the color indices $a, b=1, \cdots, 8$.

In addition, at temperatures $\mchi\lesssim \treh\lesssim\mgluino$, neutralino
decays to axinos also contribute, while at higher temperatures they are
sub-dominant. The relevant contribution is given by the decay of the
bino component:
\begin{equation}
\label{eq:Gamma-neutralino}
\Gamma(\chi_i\ra \axino + B)
= {\alpha^2_{em} C_{a\chi_i B}^2 \over 128 \pi^3} 
{m^3_{\chi_i}\over {\left(\fa/N\right)^2}} 
\(1-{\maxino^2\over m^2_{\chi_i}}\)^3.
\end{equation}
Here $\alpha_{em}$ is the electromagnetic coupling strength,
$C_{a\chi_i B} = Z_{iB}\cayy/\cos^2\theta_W$ where $Z_{iB}$ is the
bino component of the $i$th neutralino $\chi_i$ ($i=1,2,3,4$). We use
the basis $\chi_i= Z_{i1}\bino + Z_{i2}\wino +
Z_{i3}\higgsinob + Z_{i4}\higgsinot$ of the respective fermionic
partners (denoted by a tilde) of the electrically neutral gauge bosons 
$B$ and $W_3$, and the MSSM Higgs bosons $H_b$ and $H_t$.

\underline{Discussion and Results.} \hspace{0.2cm}
We have evaluated the integrals ({\ref{eq:Yscat}) and (\ref{eq:Ydec})
numerically using expressions (\ref{eq:cross-section}),
(\ref{eq:Gamma-gluino}) and (\ref{eq:Gamma-neutralino}).
The results are presented in Fig.~2 %\ref{fig:YTP-TR}
for representative values of $\fa=10^{11}\gev$ and $\msquark=\mgluino=1\tev$.
The respective contributions due to scattering as well as gluino and
neutralino decays are marked by dashed, dash-dotted and dotted lines.

\begin{figure}
\label{fig:YTP-TR}
\include{ckkr-fig1}
\caption{$\ythaxino$ as a function of $\treh$ for representative
values of $\fa=10^{11}$GeV and $\msquark = \mgluino = 1\tev$.}
\end{figure}

It is clear that at high enough $\treh$, much above $\msquark$ and
$\mgluino$, scattering processes involving such particles dominate the
axino production.  For $\treh\gg m_{\tilde q},m_{\tilde g}$,
$\langle\sigma_n v\rangle$ is almost constant and thus $Y^{\rm
scat}\simeq 2\times10^{-5}\,\mplanck\treh\times(\sum_n\sigma_n)$.  In other
words, $Y^{\rm scat}$ grows linearly as $\treh$ becomes larger.  (In
the numerical calculation we have also taken into account the running of the
strong coupling constant.  Then $Y^{\rm scat}$ is given, to a very good
approximation,  by the same formula with $\alpha_s$ replaced by
$\alpha_s(\treh)$, and it grows like $\alpha_s^3(\treh)\treh$.  Using
$\alpha_s(\treh)$ instead of its value at $M_Z$ gives a correction of up
to a factor 10. For example, $\alpha_s(M_Z)^3/\alpha_s(10^8{\rm
GeV})^3=5.75$.)
A similar result was estimated and used previously
for the saxion in Ref.~\cite{ck}.
In contrast, the decay contribution above the gluino
mass threshold, $Y^{\rm dec}\simeq 5\times10^{-4}\,(\mplanck\Gamma_{\tilde
g}/m_{\tilde g}^2)$, remains independent of $\treh$.
This is not surprising since the scattering term in Eq.~(\ref{eq:Boltzmann})
behaves like ${\rm const}\times n_in_j\sim{\rm const}\times T^6$
relative to the decay term which scales like
$\mgluino^3\ngluino\sim \mgluino^3 T^3$ plus additional suppression
$\VEV{\Gamma}\sim T^{-1}$.
In the region of very high $\treh\sim10^9\gev$,
$\ythaxino$ becomes comparable with the axino yield in thermal equilibrium
$\yeqaxino\simeq2\times10^{-3}$.  Since, as mentioned below
Eq.~(\ref{eq:Boltzmann}), we have neglected processes of
axino re-annihilation to thermal-bath particles, there appears an
artificial ``edge'' between the sloping solid curve representing $\ythaxino$
and the dashed horizontal line corresponding to the axino equilibrium
value $\yeqaxino\approx 2\times 10^{-3}$. Including axino
re-annihilation would have the effect of rounding it. We leave it as
is because we are not interested in very large values of $\treh$.

At $\treh$ roughly below the mass of the squarks and gluinos, their
thermal population starts to become strongly suppressed by the
Boltzmann factor $e^{-m/T}$, hence causing a distinct knee in the scattering
contribution in Fig.~2. %\ref{fig:YTP-TR}.
It is in this region that gluino decays
(dash-dotted line) given by Eq.~(\ref{eq:Gamma-gluino}) become
dominant, before they also become suppressed by the Boltzmann factor
due to the gluino mass.
For $\mchi\lesssim \treh\lesssim m_{\tilde q},m_{\tilde g}$,
the axino yield is well approximated by
$Y^{\rm TP}\approx Y^{\rm dec}\simeq
5\times10^{-4}(\mplanck\Gamma_{\tilde g}/\treh^2)\, e^{-m_{\tilde g}/\treh}$,
and depends sensitively on the reheating temperature.

At still lower temperatures the population of strongly interacting
sparticles becomes so tiny that at $\treh\sim\mchi$ neutralino decays
given by Eq.~(\ref{eq:Gamma-neutralino}) start playing some role, until they
too become suppressed by the Boltzmann factor. We indicate this by plotting
in Fig.~2 %\ref{fig:YTP-TR}
the contribution of the lightest
neutralino (dotted line). By comparing Eqs.~(\ref{eq:Gamma-gluino})
and~(\ref{eq:Gamma-neutralino}) we can easily estimate that the bino-like
neutralino contribution is suppressed by
$\left(\alpha_{em}/8\alpha_3\right)^2\left(\mchi/\mgluino\right)^3\sim
10^{-4}$ (with the color factor of $8$ included),
assuming the usual gaugino mass relations of the MSSM.
It is clear that the values of $\ythaxino$ in this region are so
small that, as we will see later, they will play no role in further
discussions. We therefore do not present the effect of the decay of the
heavier neutralinos.

As noted at the end of Section~\ref{sec:axino}, there are also
dimension-4 operators contributing to axino production processes.
One such operator is given by
$C_{\tilde a q\tilde q} (m_{\tilde q}/f_a)
\bar{\tilde a}\gamma_5 q{\tilde q}^\ast$, where
$C_{\tilde aq\tilde q}$ is an effective axino--quark--squark coupling
whose size is model-dependent, but which usually arises at 
the two-loop level.
It contributes to processes involving quarks in Table~1. For example,
in channel~G, in addition to the $t$-channel gluon exchange diagram
drawn in Fig.~\ref{fig:ckkr-fdiag1}, there will now be a diagram
with a squark exchange in the $s$-channel. It will contribute 
$\alphas (C_{\tilde a q\tilde q} m_{\squark}/\fa)^2(1/s)$ compared to
$\alphas^3/\fa^2$ of dimension-5 operators.  Hence it is
suppressed at high energies, $s\gg m_{\tilde q}^2$, by a factor
$(C_{\tilde aq\tilde q}/\alpha_s)^2(m_{\tilde q}^2/s)$, and it can be
non-negligible only around the mass threshold $s\sim m_{\tilde q}^2$.
This will, however, have a relatively small effect on the integration
of the Boltzmann equation, unless $T_R\sim m_{\tilde q}$.
For the DFSZ case, there is an additional contribution to the axino 
production coming from diagrams involving the Higgs supermultiplet
in the initial or final state; we neglect such part in the present study
since it is model-dependent and we expect it to be subdominant.

The sensitivity of $\ythaxino$ to the Peccei--Quinn scale
$\fa$ is presented in Fig.~3. %\ref{fig:YTP-TR-Fa}.
As is clear from Eq.~(\ref{eq:cross-section}),
the rate of axino production is inversely proportional to $\fa^2$.

\begin{figure}
\label{fig:YTP-TR-Fa}
\include{ckkr-fig2}
\caption{$\ythaxino$ as a function of $\treh$ for representative
values of $\fa$ and $\msquark=\mgluino=1\tev$.}
\end{figure}

We emphasize that axinos produced in this way are already out of
equilibrium. Their number density is very much smaller than $\ngamma$
(except  $\treh\sim10^9\gev$ and above) 
and cross sections for axino re-annihilation into other particles are
greatly suppressed. This is why in Eq.~(\ref{eq:Boltzmann}) we have neglected
such processes. Nevertheless, even though axinos never reach equilibrium, 
their number density may be large enough to give $\omegaa\sim 1$ for
large enough axino masses ($\kev$ to $\gev$ range), as we will see
later.

Before closing this section, it is worth discussing how the presence
of saxions could influence the above results. The saxions will be produced
thermally in a way analogous to the axinos but next they will
decay with a relatively short lifetime $\tau_{\rm s}$ given by
\beq
\tau_{\rm s}=2.65\times 10^{-6}\, {\rm sec}\;
\left({{\fa/N}\over{10^{11}\gev}}\cdot {{0.1}\over{\alphas}}\right)^2
\left(\frac{m_s}{1\tev}\right)^{-3}.
\eeq
This time needs to be compared with the time $\tilde t_{\rm s}=2.9\times
10^{-5}\left({1\tev}/{m_s}\right)^2$ when the saxions dominate the
energy density. If $\tau_{\rm s}>\tilde t_{\rm s}$, 
the saxion decay generates significant entropy and hence dilutes the
particle species decoupled before the decay, such as the axions and
the axinos. In other words, the present axion density is lowered and
therefore the upper bound on $\fa$  is relaxed to values well beyond
$10^{12}\gev$. Furthermore, the photon temperature now decreases more
slowly, as pointed out in Ref.~\cite{ck}.
These effects become important once the saxion mass is
greater than $5\tev$~\cite{kim91}} and $\fa$ is large.  In this study,
however, we will neglect the reheating effect due to saxion decay
since we will assume that the saxion mass is below $1\tev$. Finally, in
case the saxion mass is much smaller than the SUSY-breaking scale, as
for instance in no-scale models, then the saxion decays very late and
can play the role of a late-decaying particle~\cite{ck} instead of
generating entropy.

%%%%%%%%%%%%%%%%%%%%%%%%%%%%%%%%%%%%%%%%%%%%%%%%%%%%%%%%%%%%%%%%%%%%%%%%%%%
\section{Non-Thermal Production}
\label{sec:NTP}

As discussed in the Introduction, axinos may also be produced in decay
processes of particles which themselves are out of equilibrium,
the decaying particle being one of the 
ordinary superpartners, the gravitino or the inflaton field.
Below we will concentrate on the first possibility.

Let one of the ordinary superpartners be the LOSP and the NLSP
(next-to-lightest supersymmetric particle). (They do not have to
be the same. The case when the role of the NLSP will be taken by the
gravitino will be discussed below.)  A natural, albeit not unique,
candidate for the LOSP is the lightest neutralino. This is because its
mass is often well approximated by the bino mass parameter $\mone$
(nearly pure bino case) or by the $\mu$-parameter (higgsino limit),
neither of which grows much when evolved from the unification scale
down to the electroweak scale. It is therefore natural to expect the
neutralino to be the LOSP. Furthermore, in models employing full
unification of superpartner masses (such as the CMSSM/mSUGRA), a
mechanism of electroweak symmetry breaking by radiative corrections
typically implies $\mu^2\gg\mone^2$. As a result, the bino-like
neutralino often emerges as the lightest ordinary
superpartner~\cite{rr93,kkrw}.

Axino production from bino-like neutralino decay was analyzed in the
previous paper~\cite{ckr}.  For the sake of completeness, we
summarize here the relevant results as well as make additional points regarding
the case of a general type of neutralinos.  The process involves two
steps. The LOSPs first freeze out of thermal equilibrium
and next decay into  axinos.  Since the LOSPs are no longer in
equilibrium, the Boltzmann suppression factor in this case does not
depend on the thermal-bath temperature and has to be evaluated at the LOSP
freeze-out temperature, which is well
approximated by $\tf \simeq {\mchi/20}$.  At $T < \tf $ and for
$\Gamma_\chi \ll H$ but large enough for the decay to occur in
the radiation-dominated era, one finds that $\ychi$ is roughly given 
by~\cite{ckr}
\beq 
\ychi (T) \simeq \yeqchi (\tf)
\exp\[-\int_{T}^{\tf} \frac{dT'}{T'^3} {\mchi^2\langle
\Gamma_\chi\rangle_{T'} \over H(\mchi) }\],
\label{nchidecay:eq}
\eeq
where $\yeqchi (\tf)$ contains the Boltzmann
suppression factor evaluated at $\tf$, $\langle \Gamma_\chi \rangle_T
$ is the thermally averaged decay rate for the neutralino at
temperature $T$, and $H(\mchi)$ is given below
Eq.~(\ref{eq:Boltzmann}). While we have used here the neutralino as
the LOSP, the same mechanism would work for other ordinary
superpartners  as well.

It is clear that, in order for a two-step process to occur, the decay
width of the LOSP has to be sufficiently small to allow for the
freeze-out in the first place. Fortunately this is what usually happens
since the interactions between the LOSP and the axino are suppressed by
the large scale $f_a$.  On the other hand, the lifetime of the LOSP
must not be too large, otherwise the decay into axinos and ordinary
particles would take place too late, during or after nucleosynthesis,
and could destroy successful predictions for the abundance of light
elements. This will be discussed in more detail in the next section.
Here we note that it is truly remarkable that the relative strength of
the axino interaction, in comparison with SM interactions, is such that
the decay width falls naturally between these two limits.

It is worth mentioning that in principle the LOSP may decay into
axinos even in the absence of freeze-out.  It has also been recently
pointed out in Ref.~\cite{gkr00} that even in the case of very low reheating
temperatures $\treh$, below the LOSP freeze-out temperature, a
significant population of them will be generated during the reheating
phase. Such LOSPs would then also decay into axinos as above. The
process will be non-thermal in the sense that the decaying LOSPs will
not be (yet) in thermal equilibrium. We will not pursue this
possibility here but comment on it again when we discuss our results.

In Ref.~\cite{ckr} we considered the non-equilibrium process
\begin{equation}
\label{chitoagamma:eq}
\chi\ra\axino\gamma.
\end{equation}
This decay channel is always allowed.
The decay rate for process~(\ref{chitoagamma:eq}) can be easily
derived from Eq.~(\ref{eq:Gamma-neutralino}):
\beq
\Gamma(\chi\ra\axino\gamma) = 
{\alpha^2_{em} \cachigamma^2 \over 128 \pi^3} 
{{\mchi}^3\over {\left(\fa/N\right)^2}} \(1-{\maxino^2 \over \mchi^2}\)^3,
\label{gammachi:eq}
\eeq
where $\cachigamma=(\cayy/\cos\theta_W) Z_{11}$, with $Z_{11}$ standing for
the bino part of the lightest neutralino~\footnote{
In Eq.~(\protect\ref{gammachi:eq}) we have corrected an
overall numerical factor
with respect to Eq.~(6) of Ref.~\protect\cite{ckr}.}.

The neutralino lifetime can be written as
\beq
\tau(\chi\ra\axino\gamma)= 0.33\, {\rm sec}\, {\frac{1}{\cayy^2 Z_{11}^2}}
\left({\frac{\alpha^2_{em}}{1/128}}\right)^{-2}
\left(\frac{f_a/N}{10^{11}\gev}\right)^2
\left(\frac{100\gev}{\mchi}\right)^3 \(1-{\maxino^2 \over \mchi^2}\)^{-3}.
\label{chilife:eq}
\eeq

For large enough neutralino masses, an additional 
decay channel into axino and $Z$ opens up, 
\beq
\Gamma(\chi\ra\axino Z) = 
{\alpha^2_{em} \cachigamma^2 \over 128 \pi^3} \tan^2\theta_W
{{\mchi}^3\over {\left(\fa/N\right)^2}} \times 
{\rm PS}\left({m_Z^2 \over \mchi^2},{\maxino^2 \over \mchi^2}\right),
\label{gammachitoz:eq}
\eeq
where the phase-space factor is given by
\beq
{\rm PS}(x,y) = \sqrt{1-2 (x+y) + (x-y)^2} \left[\(1-y\)^2 
- {x\over 2}\(1- 6\sqrt{y}+ y\) - {x^2\over 2} \right].
\eeq

Note that this channel is always subdominant relative to
$\chi\ra\axino\gamma$ because of both the phase-space suppression and
the additional factor of $\tan^2\theta_W$. As a result, even at
$\mchi\gg \mz,\maxino$, $\tau(\chi\ra\axino Z)\simeq 
3.35\, \tau(\chi\ra\axino\gamma)$. 
It is also clear that the neutralino lifetime rapidly decreases with its
mass ($\sim1/\mchi^3$).  On the other hand, if the neutralino is not
mostly a bino, its decay will be suppressed by the $Z_{11}$-factor
in $\cachigamma$. 

Other decay channels are the decay into axino and Standard Model
fermion pairs through virtual photon or $Z$, but they are negligible
with respect to the previous ones. We will discuss them later since,
for a low neutralino mass, \ie\ long lifetime, they can, even if
subdominant, produce dangerous hadronic showers during and after
nucleosynthesis.

In the DFSZ type of models, there exists an additional
Higgs--higgsino--axino coupling, usually related to the MSSM $\mu$-term
in simplest realizations. For example, by using the fields $S_1$ and
$S_2$, which break the PQ symmetry in the superpotential given
by~(\ref{kim83}), one can generate the MSSM Higgs mass term through
the renormalizable interactions 
\beq 
W_\mu = \lambda_i S_i H_u H_d
\eeq 
for $i=1,2$, with a very small $\lambda_i \simeq
m_W/f_a$ or through a non-renormalizable interaction like 
\beq W_\mu
= \lambda^\prime_i {S^2_i H_u H_d \over \mplanck}.  
\eeq
 
After the PQ symmetry breaking, such superpotentials not only generate
the MSSM $\mu$-term~\cite{susyaxion2,muterm}, but in general also give rise to a
Higgs--higgsino--axino coupling of order $\mu/\fa $. In the
non-renormalizable case a four-fermion coupling of order $\mu/\fa^2$
is generated.  This coupling can be neglected since it does not
contribute to the neutralino decay but the other can if the higgsino
component of the neutralino is non-negligible.  Note that after
electroweak symmetry breaking such couplings produce a mixing between
the axino and the higgsinos but in general these are of the order of $
\mu v/\fa$, and therefore much less than $\mu$, so that we can
continue to consider the axino as an approximate mass eigenstate which
is nearly decoupled from the MSSM.

An example of such an effective MSSM+axino model has been recently analyzed
in Ref.~\cite{martinaxino} for an effective potential of the type
\beq
W^{\rm eff}_\mu = \mu \(1 + {\epsilon\over v} \Phi \) H_u H_d,
\eeq
where $\epsilon \simeq v/\fa \simeq 10^{-8}$
and $v$ is the Higgs VEV of the order of the weak scale. 
One needs to perform a full diagonalization of the
$5\times5$ neutralino+axino mass matrix.
Since in this case the higgsinos
couple more strongly (or rather, less weakly) to the axino,
the non-negligible mixing between axino and higgsino introduces
additional decay channels, \eg\ the one with intermediate virtual
right-handed sleptons, that are absent in the KSVZ case.

By rescaling the results of Fig.~2 of~\cite{martinaxino} to our central
value $\fa = 10^{11} \gev$, 
one finds that for low bino mass, $\mchi \leq 120 \gev $, $\maxino =
50 \gev$, and specific choices of other
supersymmetric parameters, in particular masses of the right-handed
sleptons degenerate and larger than $\mchi $,
the decay time of the neutralino into axino
and lepton or quarks pairs is of the order of
\beq 
\tau (\chi
\rightarrow \tilde a l\bar l \,\, \mbox{and}\,\, \tilde a q\bar q)
\simeq 0.02\, {\rm sec};
\eeq 
it decreases at higher masses because of the dependence on $\mchi^3$
and of the
opening of new channels ($b\bar b$ and $ W W^*$), down to $ 3 \times
10^{-6}\, {\rm sec}$ at $\mchi = 200 \gev$. We then see that even in
the DFSZ 
case the lifetime is of the right order of magnitude to allow the
freeze-out and decay process since $ H(\mchi) \simeq 2 \times 10^7\, 
{\rm sec}^{-1} (\mchi/100 \gev)^2 $.  

Of course this result strongly depends on the type of DFSZ model
considered: for example, the coupling $\epsilon $ between 
the axino and the Higgses can be suppressed or enhanced by 
mixing angles in the PQ sector since the axion multiplet is in
general a combination of different multiplets ($S_1$ and $S_2$ 
in the above example, compare Eq.~(\ref{kim83})). Such angles 
are assumed to be of order 1 in the simple estimate
above.

We therefore conclude that DFSZ-type models have to be
analyzed case by case, but in general we expect them to give  
an even more efficient implementation  of the non-thermal production 
through LOSP decay. A scenario of this type has been recently
studied in detail in~\cite{ckl00} with the conclusion that an 
axino (or better in this case flatino) LSP and CDM candidate is 
still possible and that its population is produced naturally after 
a phase of thermal inflation, with very low reheat temperature.

While in the discussion of the non-thermal production of axinos, we have
concentrated on the neutralino as, in some sense, the most natural
choice for a parent LOSP, we reiterate that in principle one could
also consider other choices for the LOSP, including charged particles,
which we will not do here.

Finally, we note for completeness that another way 
of producing axinos non-thermally is to consider gravitino
decays~\cite{ay00}. In this case it is the gravitino, and not the
LOSP, that is the NLSP. Such axinos are
produced very late, at $10^8\, {\rm sec}$. As we have stated in the
Introduction, this channel may solve the gravitino
problem~\cite{ay00}. We will come back to this point in the next Section.
Furthermore, axinos may be produced during (p)reheating in the decay
of the inflaton field. Such processes are strongly model-dependent and
we will not consider them here.

%%%%%%%%%%%%%%%%%%%%%%%%%%%%%%%%%%%%%%%%%%%%%%%%%%%%%%%%%%%%%%%%%%%%%%%%%%%
\section{Constraints}
\label{constraints:sec}

Several non-trivial conditions have to be satisfied in order for
axinos to be a viable CDM candidate. First, we will expect their relic
abundance to be large enough, $\abunda\simeq0.2$. This obvious
condition will have a strong impact on other bounds. Next, the
axinos generated through both TP and NTP will in most cases be
initially relativistic. We will therefore require that they become
non-relativistic, or cold, much before the era of matter
dominance.  Furthermore, since NTP axinos will be produced near the
time of BBN, we will require that they do not contribute too much
relativistic energy density to radiation during BBN. Finally,
associated decay products of axino production will often result in
electromagnetic and hadronic showers which, if too large, would cause
too much destruction of light elements. In deriving all of these
conditions, except for the first one, the lifetime of the parent LOSP
will be of crucial importance.

First, from now on we will assume that the axinos give a dominant
contribution to the matter density at the present time. This can be
expressed as
\beq
\abunda = \maxino\yaxino\;{s(T_{\rm now})\over \rhocrit/h^2} \simeq 0.2,
\label{abundacond:eq}
\eeq 
where $s$ is given under Eq.~(\ref{ythaxino:eq}) and we have assumed 
no significant entropy production.
Since $\rhocrit/h^2=0.8\times10^{-46}\gev^4$ and $\gsstar(T_{\rm now})=3.91$,
this can be re-expressed as 
\beq 
\maxino \yaxino \simeq 0.72 \ev \(\abunda \over 0.2\),
\label{eq:m-yaxino}
\eeq
which readily applies to both TP and NTP relics.

We note in passing that, for the initial population of axinos, the
yield at decoupling is approximately
$\yaxino\simeq\yeqaxino\simeq2\times10^{-3}$, which gives 
\beq 
\maxino \simeq 0.36 \kev \(\abunda \over 0.2\).
\label{eq:warmaxino}
\eeq
This is an updated value for the Rajagopal--Turner--Wilczek
bound~(\ref{eq:rtw:warmaxino}). 

For a fixed value of $\abunda$, lower values of $\yaxino$ give
correspondingly larger $\maxino$. For example, the turn-over region of
$\yaxino\simeq10^{-10}$ in Fig.~2, where gluino decay becomes more
important than scattering, gives $\maxino \simeq 7.2 \gev \(\abunda/
0.2\)$.

\underline{Cold Axinos} \hspace{0.2cm}
Next, we want to determine the temperature of the Universe at which
the axinos will become non-relativistic. 
In nearly all cases axinos  are initially relativistic
and, due to expansion, will 
become non-relativistic at some later epoch,
which depends on their mass and  production mechanism.
As mentioned in the Introduction, in the case of thermal production,
even though the axinos are not in thermal equilibrium, they are
produced in kinetic equilibrium with the thermal-bath. Their momenta
therefore have a thermal spectrum, derived from the scattering
particles in the plasma.  The axinos
become non-relativistic when the thermal-bath temperature reaches the
axino mass:
\beq
\Tnr\simeq \maxino, 
\label{eq:tnrtp}
\eeq
even though  they are not in thermal equilibrium. 

In the case of non-thermal production, the momentum of the axinos
depends strongly on the production mechanism. Here we will consider,
as before, only the production through out-of-equilibrium neutralinos.
In that case, axinos will be produced basically monochromatically, all
with the same energy, roughly given by $\mchi/2$, unless they are
nearly mass-degenerate with the neutralinos. This is so because the
neutralinos, when they decay, are themselves already non-relativistic.
Thus the axinos will normally become non-relativistic only at later
times, through momentum red-shift.  However, in this case the simple
estimate~(\ref{eq:tnrtp}) used above for $\Tnr$ does not hold, since
the axinos are not in kinetic equilibrium with the thermal-bath.

First we need to determine the temperature $\Tdec$ corresponding to
the time of neutralino decay $\tdec$. We will make the approximation that
most axinos are produced in neutralino decays at the time equal to the
neutralino lifetime $\tau_\chi=1/\Gamma_\chi$, $\tdec\simeq\tau_\chi$
(a sudden-decay approximation), which is a very reasonable one.

Because not all neutralinos decay at the same time as assumed in the
sudden-decay approximation, at a given time the momenta of produced
axinos will actually have a distribution that will be limited from above by
$m_\chi/2$.  Let us find this momentum distribution $f_{\tilde
a}(t_0,p)$ of axinos coming from neutralino decays.  When the axino
was produced by the decay of neutralino at time $t$, it had the
momentum $p=m_\chi/2$, and would be next red-shifted while the axino was
relativistic.  Since the scale factor $R$ grows as $R\propto t^{1/2}$
during the RD epoch, at a later time $t_0$, the axino produced at a
time $t$ would have the momentum
\begin{equation}
p(t,t_0) = \frac{m_\chi}{2} \left(\frac{R(t)}{R(t_0)}\right)
  = \frac{m_\chi}{2} \left(\frac{t}{t_0}\right)^{1/2}.
\end{equation}
By counting the number of produced axinos during the time interval $dt$ at
$t$, we get
\begin{equation}
N_\chi\frac{dt}{\tau_\chi}=f_{\tilde a}(t_0,p)dp,
\end{equation}
where $N_\chi(t)=N_{\chi0}e^{-t/\tau_\chi}$ is the number of neutralinos at $t$
which is obtained from the decay equation $dN_\chi/dt=(1/\tau_\chi)N_\chi$.
Thus,
\begin{equation}
f_{\tilde a}(t_0,p) = \frac{N_\chi}{\tau_\chi}\frac{dt}{dp}
= N_{\chi0} \frac{2p}{p(\tau_\chi,t_0)^2}
e^{-\left(p/p(\tau_\chi,t_0)\right)^2}.
\end{equation}
From this, we can find the total energy of axinos
\begin{equation}
E_{\tilde a}(t_0) = \int_0^{m_\chi/2} pf_{\tilde a}(t_0,p)dp
= N_0p(\tau_\chi,t_0)
\left[\frac{\sqrt{\pi}}{2}{\rm
Erf}\left[\left(\frac{t_0}{\tau_\chi}\right)^{1/2}\right]
-\left(\frac{t_0}{\tau_\chi}\right)^{1/2}e^{-t_0/\tau_\chi} \right].
\end{equation}
For $t_0\gg\tau_\chi$, the energy density of axinos becomes
\begin{equation}
\rho_{\tilde a} = \frac{E_{\tilde a}(t_0)}{R(t_0)^3}
= \frac{\sqrt{\pi}}{2}\;\frac{m_\chi}{2}
\left(\frac{\tau_\chi}{t_0}\right)^{1/2} \frac{N_{\chi0}}{R(t_0)^3}.
\end{equation}
This differs from the sudden-decay approximation only by a factor
$\sqrt{\pi}/2\approx0.89$.

Using the sudden-decay approximation we obtain
\bea
\Tdec &=&  \(\frac{90\; \Gamma_\chi^2 \mpl^2}{4\pi^2 \gstar(\Tdec)}\)^{1/4}\\
      &=& 1.08\times 10^{9}\gev\( {{\Gamma_\chi}\over{\gev}} \)^{1/2} 
       \simeq  0.9 \mev\( {{\rm sec}\over{\tau_\chi}}\)^{1/2},
\label{eq:Tdec}
\eea
where we have taken $\gstar(\Tdec)\simeq10$.

Axino momenta will red-shift with temperature as
\beq
p(T) = {\mchi\over 2} 
\left({\gsstar (T)\over \gsstar (\Tdec)}\right)^{1/3} {T\over \Tdec}.
\label{eq:paxino}
\eeq
as long as the axinos remain relativistic.
(The factor involving $\gsstar $
at different times accounts for a possible reheating of the
thermal-bath when some species become non-relativistic.)
Axinos become non-relativistic at the epoch when
$p(\Tnr) \simeq \maxino$, which gives
\bea
\Tnr  &=& 2\, {\maxino\over \mchi}
\left({\gsstar(\Tdec)\over \gsstar(\Tnr)}\right)^{1/3} \Tdec \\
   &=& 2.7 \times 10^{-5}\maxino \({100 \gev \over \mchi}\) 
    \({\Tdec\over 1 \mev}\),
\label{eq:Tnr}
\eea
where we have taken $\gsstar(\Tdec)\simeq\gstar(\Tdec)
\simeq 10$ and $\gsstar(\Tnr) = 3.91$.

Using an explicit expression for $\Tdec$ (Eq.~(\ref{eq:Tdec})) 
then leads to
\bea
\Tnr & =& 
2\, {\maxino\over \mchi}
\left({\gsstar(\Tdec)\over \gsstar(\Tnr)}\right)^{1/3}
\(\frac{90\Gamma_\chi^2 \mpl^2}{\pi^2 \gstar(\Tdec)}\)^{1/4}\\
& =& 53.8 \kev 
\left({\gsstar(\Tdec)\over \gsstar(\Tnr)}\right)^{1/3}
{\cayy Z_{11} \over \gstar^{1/4} (\Tdec)}
\left({\maxino\over 1 \gev}\right)
\left( {\mchi\over 100 \gev} \right)^{1/2}
\left({10^{11} \gev\over f_a/N}\right).
\eea
For $\Tdec< 200\mev$, $\gsstar(\Tdec)\simeq\gstar(\Tdec)
\simeq10$ 
and for 
$T< 500\kev$, $\gsstar(T)\simeq 3.91$, which finally leads to
\beq
\Tnr  = 4.2\times 10^{-5}\maxino \cayy Z_{11}
\left( {\mchi\over 100 \gev} \right)^{1/2}
\left({10^{11} \gev\over f_a/N}\right).
\label{eq:tnrfinal}
\eeq

This epoch has to be compared with the matter-radiation equality
epoch. Assuming that axinos constitute the largest part of the
matter density, $\rho_{\rm matter}(\Teq)=\rhoaxino(\Teq)$, and since
\beq
\rhoaxino= \maxino \yaxino s(T) = 
\maxino \yaxino {2\pi^2\over 45} \gsstar (T)\; T^3,
\label{eq:rhoaxino}
\eeq
the temperature at matter-radiation equality is given by
\beq
\Teq = {4 \gsstar (\Teq)\over 3 \gstar(\Teq)}  
\maxino \yaxino.
\label{eq:tequality}
\eeq
Using Eq.~(\ref{eq:m-yaxino}) and assuming $\gsstar (\Teq)= 3.91$ and
$\gstar(\Teq)= 3.36$ 
allows us
to express Eq.~(\ref{eq:tequality}) as
\beq
\Teq = 1.1 \ev \(\abunda \over 0.2\),
\label{eq:tequalityfinal}
\eeq
which holds for both thermal and non-thermal production.

In the former case, by comparing with Eq.~(\ref{eq:tnrtp}), one can
easily see that $\Tnr>\Teq$ is satisfied for any interesting range of
$\maxino$.  In the case of NTP, in Eq.~(\ref{eq:tequality}) we make a
substitution $\yaxino=\ychi (\tf) $, the neutralino yield at
freeze-out.  The condition $\Tnr\gg\Teq$ is satisfied for 
\bea
\maxino &\gg & 41 \kev \( {\mchi\over 100 \gev} \) 
\( {1 \mev\over \Tdec} \) \(\abunda \over 0.2\)
\label{eq:relaxinobbnone} \\
&\gg & 27 \kev {1\over \cayy Z_{11}} \( {100 \gev \over \mchi} \)^{1/2} 
\left({ f_a/N\over 10^{11} \gev}\right),
\(\abunda \over 0.2\)
\eea
where the $\Tdec$ dependence on the neutralino lifetime is given 
in Eq.~(\ref{eq:Tdec}).
If axinos were lighter than the bound~(\ref{eq:relaxinobbnone}), 
then the point of radiation-matter equality would be shifted
to a later time around $\Tnr $. Note that in this case the axino
would not constitute cold, but warm or hot dark matter.
We will see, however,  that in the NTP case discussed here
other constraints would require the axino 
mass to be larger than the above bound, so that we can discard this
possibility.

\underline{BBN constraint on NTP axinos} \hspace{0.2cm}
In the case of non-thermal production, most axinos will be produced
only shortly before nucleosynthesis and, being still
relativistic, may dump too much to the energy density during
the formation of light elements. The axino energy density now reads

\bea
\rhoaxino(T) &=& \yaxino s(T)p(T) \nonumber\\
&=& \left(\frac{2\pi^2}{45}\right)
    \left(\frac{\gsstar (T)^{4/3}}{\gsstar(\Tdec)^{1/3}}\right)
    \left(\frac{\mchi\yaxino}{2\Tdec}\right) T^4 \nonumber\\
&=& 1.6\times 10^{-7}
\left(\frac{\gsstar (T)^{4/3}}{\gsstar(\Tdec)^{1/3}}\right)
\left(\frac{\mchi}{\maxino}\right)
\left(\frac{1\mev}{\Tdec}\right)
\left(\frac{\abunda}{0.2}\right) T^4,
\label{eq:relbbnpart}
\eea
where Eqs.~(\ref{eq:m-yaxino}), (\ref{eq:paxino}), 
and an expression for the entropy density $s$ given under
Eq.~(\ref{ythaxino:eq}) have been used.
In order not to affect the Universe's expansion during BBN, the axino
contribution to the energy density should satisfy
\beq
{\rhoaxino\over \rhonu} \leq \deltannu,
\label{eq:relbbn}
\eeq
where $\rhonu=({\pi^2/ 30})(7/4) T^4$ is the energy density of one
neutrino species. Agreement with observations of light elements 
requires~\cite{bbnbound} 
$\deltannu = 0.2$ to $1$. Using Eq.~(\ref{eq:Tdec}) leads,
after some simple algebra, to 
\bea
\maxino &\gsim & 274 \kev {1\over \deltannu} \( {\mchi\over 100 \gev} \)
 \( {1 \mev\over \Tdec} \) \(\abunda \over 0.2\)
\label{eq:relaxinobbn}\\
&\gsim & 181  \kev {1\over \deltannu} {1\over \cayy Z_{11}} 
\( {100 \gev \over \mchi} \)^{1/2} \left({ f_a/N\over 10^{11} \gev}\right)
\(\abunda \over 0.2\),
\eea
where we have chosen the most conservative case $\gsstar(T) =
\gsstar(\Tdec) \simeq 10$. (For $\gsstar(T) = 3.91 $ the bounds become
respectively $ 78\kev $ and $55\kev $.) As before, $\Tdec$ depends on
the neutralino lifetime via Eq.~(\ref{eq:Tdec}).

\underline{Showers} \hspace{0.2cm} 
In the NTP case, if neutralino decays take place during or after
BBN, produced bosons may lead to a significant depletion of
primordial elements~\cite{kt}. One often applies the crude constraint
that the lifetime be less than about $1\, {\rm sec}$. This in our case
would provide a lower bound on $\mchi$. A detailed
analysis~\cite{bbnbound} provides limits on the abundance of the
decaying particle versus its lifetime. (See, \eg, Fig.~3 of
Ref.~\cite{bbnbound}.)

First, the photons produced in reaction~(\ref{chitoagamma:eq}) carry a
large amount of energy, roughly $\mchi/2$. If the decay takes place
before BBN, the photon will rapidly thermalize via multiple
scatterings from background electrons and positrons
($\gamma+e\ra\gamma+\gamma+e$)~\cite{dkt,kmn}. The process will be
particularly efficient at plasma temperatures above $1\mev$, which is
the threshold for background $e\bar e$ pair annihilation, and
which, incidentally, coincides with a time of about $1\, {\rm sec}$. But a
closer examination shows that also scattering with the high-energy
tail of the CMBR thermalize photons very efficiently. As a result, the decay
lifetime into photons can be as large as $10^4\, {\rm sec}$. By comparing this
with Eq.~(\ref{chilife:eq}) we find that, in the gaugino regime, this
can be easily satisfied for $\mchi<\mz$. It is only in a
nearly pure higgsino case and for a mass of tens of $\gev$ that the bound
would become constraining. We are not interested in such light
higgsinos for other reasons, as will be explained later.

A much more stringent constraint comes from considering hadronic
showers from $q\bar q$-pairs. These will be produced through a virtual
photon and $Z$ exchange, and, above the kinematic threshold for
$\chi\ra\axino Z$, also through the exchange of a real $Z$-boson.
We will now discuss this in some detail.

By comparing with Fig.~3 of Ref.~\cite{bbnbound},
we can see that the bound on hadronic showers can be written as
\beq
{\mchi \nchi \over \ngamma}\times BR(\chi\ra q\bar q) < F(\tauchi),
\label{bbnhadronic:eq}
\eeq
where $BR(\chi\ra q\bar q)$ is the branching ratio of neutralinos into
axinos plus  $q\bar q$-pairs and $F(\tauchi)$ can be read out from Fig.~3 of
Ref.~\cite{bbnbound}. 

Since $\yaxino\simeq\ychi (\tf)$, and remembering that
\beq
{\nchi\over\ngamma} = {\pi^4 \gsstar \over 45 \zeta(3) } \ychi\simeq
7.04\, \ychi,
\label{nchivsychi:eq}
\eeq
one can express the condition~(\ref{bbnhadronic:eq}) as
\beq
{\maxino\over\mchi} > 5.07 \( { 10^{-9}\gev \over F(\tauchi)} \) 
\(\abunda \over 0.2\) \times BR(\chi\ra q\bar q).
\label{condhadronic:eq}
\eeq

As stated above, one can write 
\beq
BR(\chi\ra q\bar q) = BR(\chi\ra \gamma^{\ast} \ra q\bar q) + 
BR(\chi\ra Z/Z^{\ast} \ra q\bar q) + {\rm interference\ term}.
\label{bratio:eq}
\eeq 

Assuming $\maxino\ll\mchi,\mz$, in the region $\mchi<\mz$, we find
that $BR(\chi\ra \gamma^{\ast} \ra q\bar q) $ dominates and (summing
over all light quark species up to the bottom) is of the order of
$0.03$--$0.04$, slowly increasing with $\mchi$. Above the $Z$ threshold
the intermediate $Z$ channel also becomes sizeable, and the branching
ratio grows faster with $\mchi$, up to $\simeq 0.06$ at $\mchi = 150 \gev$.
The interference part is always negligible.

We have computed $BR(\chi\ra q\bar q) $ both below and above the
threshold for the process $\chi\ra \axino Z$ for negligible axino mass
and used it in our numerical analysis.  Our results are presented in
Fig.~4 in the bino case $Z_{11} \cayy \simeq 1$ and
for $\fa/N=10^{11}\gev$. For the general case, one can still read out
the lower bound on the ratio $\maxino/\mchi$ from Fig.~4 by
replacing $\mchi$ with $(\cayy Z_{11})^{2/3} (10^{11}\gev/\fa/N)^{2/3}
\mchi $.  In fact the main dependence on $\mchi$ in
Eq.~(\ref{condhadronic:eq}) comes from the factor $\mchi$ and the
neutralino lifetime in $F(\tau_\chi)$, while $BR(\chi\ra q\bar q) $
only slowly increases with $\mchi$.

\begin{figure}
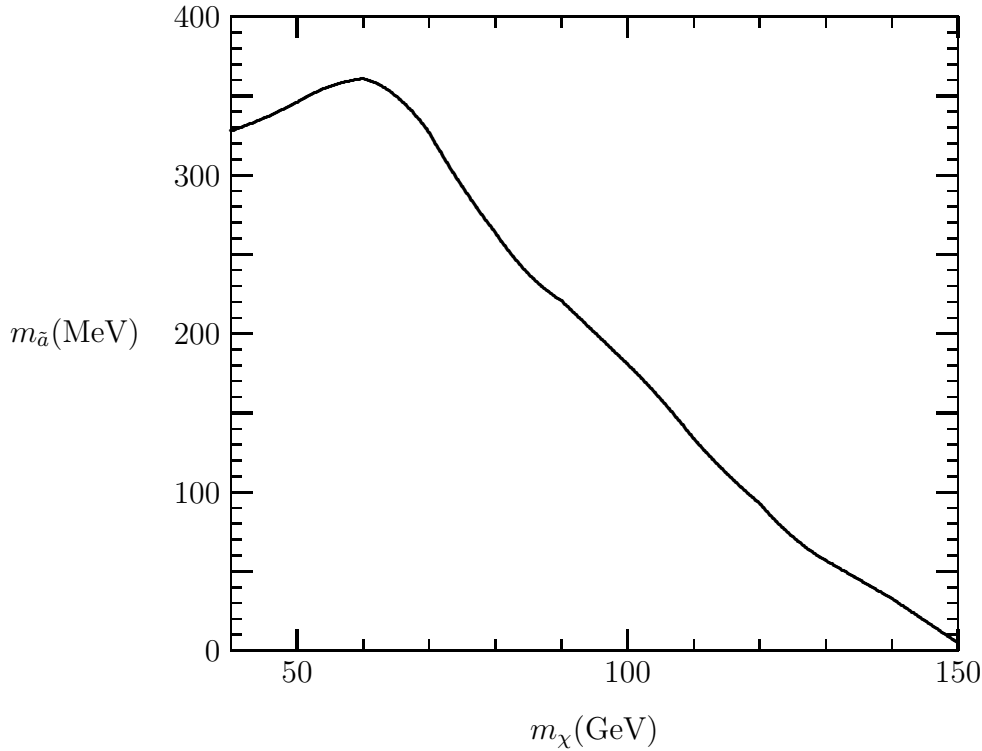

\label{fig:ma-bound}
\include{ckkr-fig3}
\caption{Lower bound on the axino mass from considering hadronic
showers according to the condition (\ref{condhadronic:eq}), for 
$\cayy Z_{11} = 1 $ and $\fa/N = 10^{11}\gev$. The bound
disappears for $\mchi = 150 \gev$ when the lifetime drops
below $0.1\, {\rm sec}$.}
\end{figure}

These bounds are clearly much more stringent than those
in~(\ref{eq:relaxinobbnone}) and~(\ref{eq:relaxinobbn}), but they are,
at the same time, strongly sensitive to the neutralino mass and
composition. Notice that, in the region of low $\mchi$, the bound on
$\maxino$ gradually strengthens with increasing $\mchi$ (and therefore
decreasing neutralino lifetime) because, for lifetimes of order
$2$--$20\, {\rm sec}$, the function $F(\tau_\chi) $ is shallow and the
branching ratio and $\chi $ both increase. However, as the lifetime
drops below $1\, {\rm sec}$, $F(\tau_\chi) $ increases steeply and so
the $\maxino$ bound decreases almost linearly in $\mchi$ before
disappearing altogether for $\mchi\gsim150\gev$ (in the bino case)
when $\tauchi\lsim0.1\, {\rm sec}$.

If the higgsino component of the decaying neutralino increases, so
does the lifetime. There are two points to note here. One is that the
neutralino yield now becomes much smaller, owing to co-annihilation with
the next to lightest neutralino and lightest chargino until
$\mchi\gsim500\gev$, when the lifetime will be suppressed
again. Furthermore, in the DFSZ model, new channels are present,
as we discussed in the previous section, for which  a typical lifetime 
will be very much smaller than $1 \, {\rm sec}$~\cite{martinaxino}. 
Only in the KSVZ model in the higgsino-like case, in a relatively small
region $60\lsim\mchi\lsim 150 \gev$ (the lower number being the current
rough experimental bound), will there exist some restriction on the
combination of neutralino mass and higgsino purity, which would further
depend on the neutralino yield at freeze-out.  Thus we conclude that
overall we find no significant restriction on the neutralino mass from
the lifetime constraint.

In summary, a lower bound $\maxino\gsim{\cal O}(300\kev)$ arises from
requiring either the axinos to be cold at the time of matter dominance
or that they do not contribute too much to the relativistic energy
density during BBN. The constraint from hadronic destruction of light
elements can be as strong as $\maxino\gsim 360 \mev$ (in the
relatively light bino case), but it is highly model-dependent and
disappears for larger $\mchi$.

%%%%%%%%%%%%%%%%%%%%%%%%%%%%%%%%%%%%%%%%%%%%%%%%%%%%%%%%%%%%%%%%%%%%%%%%%%%
\section{Thermal vs. Non-Thermal Production}
\label{sec:TPvsNTP}

We now proceed to comparing the relic abundance of axinos produced
non-thermally in neutralino decays with the thermal production case
analyzed in Section~\ref{sec:TP}.  Clearly, in the TP case the axino yield is
primarily determined by the reheating temperature (for a fixed $\fa$).
For large enough $\treh$ ($\treh\gg\mgluino,\msquark$), it is
proportional to $\treh/\fa^2$. (Compare Figs.~2 and~3.)  For the
reheating temperature below the mass threshold of strongly interacting
sparticles, it becomes Boltzmann suppressed by the factor
$\sim\exp^{-m_{\tilde g}/\treh}$ (or $\sim\exp^{-m_{\tilde q}/\treh}$).

If the axino is the LSP, the present fraction of the
axino energy density to the critical density is given by
Eq.~(\ref{eq:m-yaxino}). This relation allows us to redisplay the
results of Fig.~2 in the plane ($\maxino, \treh$).  For thermally
regenerated axinos, Eq.~(\ref{eq:m-yaxino}) gives an upper bound on
$\treh$ as a function of the axino mass from the requirement
$\omegaatp h^2\lesssim1$.
This bound is depicted by a thick solid line in Fig.~5. %\ref{fig:TR-maxino}.

\begin{figure}
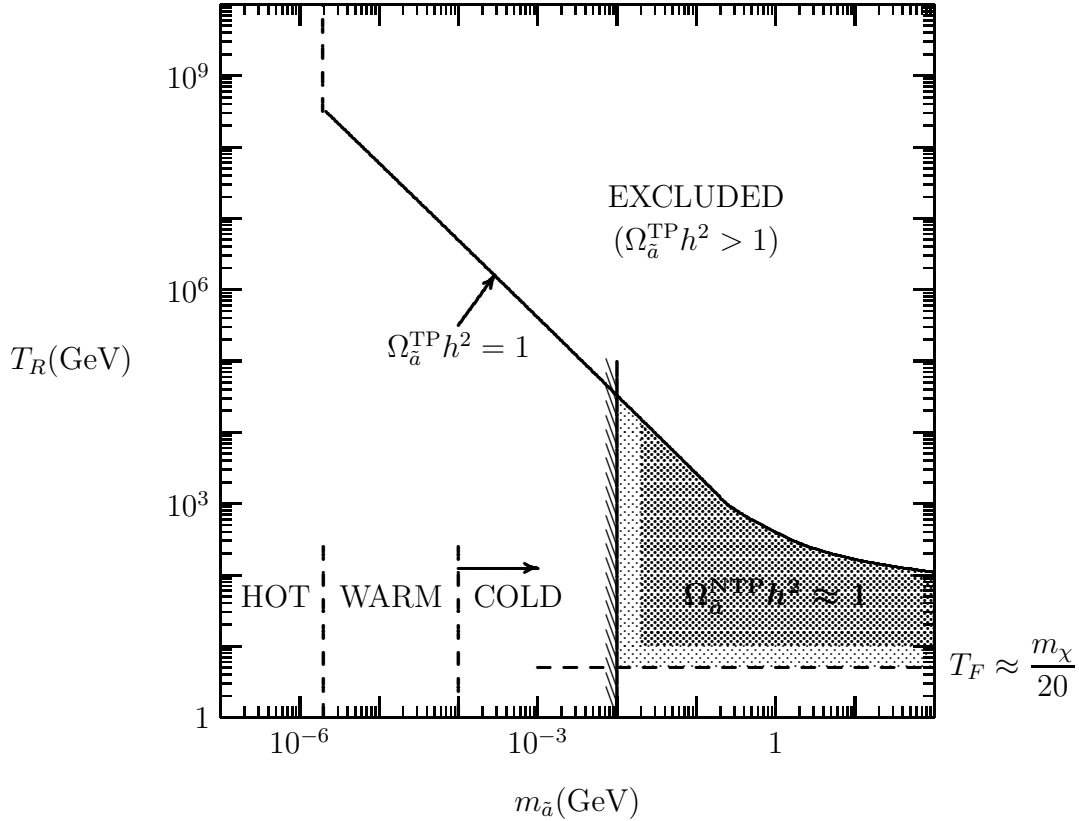

\label{fig:TR-maxino}
\include{ckkr-fig4}
\caption{
The solid line gives the upper bound from thermal production
on the reheating temperature as a function of the axino mass.  The
dark region is the region where non-thermal production can give
cosmologically interesting results ($\Omega^{\rm NTP}_{\axino}
h^2\simeq1$) as explained in the text.  
We assume a bino-like neutralino with $\mchi=100\gev$
and $\fa=10^{11}\gev$. The region of $\treh\gsim\tf$ is somewhat
uncertain and is shown with light-grey color. A sizeable
abundance of neutralinos (and therefore axinos) is expected also for
$\treh\lsim\tf$~\protect\cite{gkr00} but has not been
calculated. The vertical light-grey band indicates that a low range of
$\maxino$ corresponds to allowing SM superpartner masses in the
multi-TeV range, as discussed in the text.
The division of hot, warm and cold dark matter as a function of 
the axino mass shown
in the lower left part is for axinos from non-thermal production.}
\end{figure}

A digression is in order here. An expert reader will have noticed
that, while in the axino case, the bound $\abunda\lsim1$ gives
$\treh\sim 1/\maxino$, in the gravitino case the analogous bound gives
$\treh\sim\mgravitino$. (See, \eg, Fig.~1 in Ref.~\cite{mmy}.) This
difference is caused by the fact that the crucial effective Lagrangian
dimension-5 operator, which is responsible for the bound, is
proportional to $1/\mgravitino$ in the gravitino case, but exhibits no
$\maxino$ dependence in the case studied here. (Compare Eq.~(2) or
Ref.~\cite{mmy} with Eq.~(\ref{axionints:eq}).) 

Since axinos couple like $1/\fa$, one would naively expect that
$\treh$ for which TP becomes important (or when too much relic
abundance is generated) would be just $\fa^2/\mplanck^2$ of that for
gravitinos and thus hopelessly low, $\treh\lsim10^{-7}\mbox{GeV}$. 
This is, however, not the case, since the gravitino production is
dominated by the goldstino component, whose interaction is 
suppressed by the supersymmetry-breaking scale $M_S$, rather than the Planck 
scale: for example, the coupling to the gluino is 
$\propto\mgluino/M_S^2$. Thus $\treh$ at which
TP becomes significant will be of order $10^3$--$10^4\gev$, which is on the
low side but still acceptable.

In the NTP case, the yield of axinos is
just the same as that of the decaying neutralinos. This leads to the
following simple relation~\cite{ckr}
\begin{equation}
\abunda = {\maxino\over\mchi}\, \abundchi.
\label{abunaavsabundchi:eq}
\end{equation}
We remind the reader that $\abundchi$ stands for the abundance that the
neutralinos would have had today, had they not decayed into axinos.  It
is related to the neutralino yield through an analogue of
Eq.~(\ref{eq:m-yaxino}) (which actually applies to any stable massive
species) and is determined by the effective cross section of
neutralino pair-annihilation (as well as co-annihilation) into
ordinary particles.  When the rate of this process becomes less than
the rate of the expansion of the Universe, the neutralinos freeze out.
Typically this happens at freeze-out temperatures of
$\tf\simeq\mchi/20$.

In contrast to the TP case, the NTP axino yield will also be
independent of the reheating temperature (so long as
$\treh\gg\tf$). In order to be able to compare the two production
mechanisms, we will therefore fix the neutralino mass at some typical
value. Furthermore we will map out a cosmologically interesting range
of axino masses for which $\omegaantp\sim1$.

Our results are presented in Fig.~5 in the case of a nearly pure bino.
We also fix $\mchi=100\gev$ and $\fa=10^{11}\gev$. The dark region is
derived in the following way. It is well known that $\abundchi$, the
relic abundance of neutralinos, can take a wide range of values
spanning several orders of magnitude. In the framework of the MSSM,
which we have adopted, global scans give $\abundchi\lsim10^4$ in the
bino region at $\mchi\lsim100\gev$.
(This limit decreases roughly linearly (on a log-log scale) down to
$\sim10^3$ at $\mchi\simeq400\gev$.)
For $\mchi=100\gev$, by using Eq.~(\ref{abunaavsabundchi:eq})
we find that the expectation $\Omega^{\rm NTP}_{\axino} h^2\simeq1$ gives 
\begin{equation}
\label{ntpaxinorange:eq}
10\mev\lsim\maxino\lsim\mchi.
\end{equation}
We note, however, that the upper bound $\abundchi\lsim10^4$ comes from
allowing very large $\msusy$ (\ie\ sfermion and heavy Higgs masses) in
the range of tens of $\tev$. Restricting all SUSY mass parameter below
about $1\tev$ reduces $\abundchi$ below $10^2$ and, accordingly,
increases the lower bound $\maxino\gsim1\gev$. For the sake of
generality, in Fig.~5 we have kept the much more generous
bound~(\ref{ntpaxinorange:eq}) but we marked a low range of
$\maxino$ with a light grey band to indicate the above point. 

Likewise, for reheating temperatures just above $\tf$, standard
estimates of $\abundchi$ become questionable. We have therefore
indicated this range of $\treh$ with again light grey color.  It has
also been recently pointed out in Ref.~\cite{gkr00} that a significant
population of LOSPs will be generated during the reheating phase even
at $\treh$ below the LOSP freeze-out temperature.  Such LOSPs would
then also decay into axinos as above. We have not considered such
cases in our analysis and accordingly left the region $\treh<\tf$
blank, even though in principle we would expect some sizeable range of
$\abunda$ there.

We can see that, for large $\treh$, the TP mechanism is more important
than the NTP one, as expected.  Note also that in the TP case the
cosmologically favored region ($0.2\lsim\abunda\lsim0.4$) would form a
very narrow strip (not indicated in Fig.~5) just below the
$\omegaatp=1$ boundary.  In contrast, the NTP mechanism can give the
cosmologically interesting range of axino's relic abundance for a
relatively wide range of $\maxino$ so long as $\treh\lsim 5\times
10^4\gev$. Perhaps in this sense, the NTP mechanism can be considered
as somewhat more robust.

We have also marked in Fig.~5 some of the bounds discussed in
Section~\ref{constraints:sec}. They are normally not as restrictive as
the shaded region in the Figure. For example, the ranges of $\maxino$ where
non-thermally produced axinos would be hot/warm/cold dark matter are denoted,
following Eqs.~(\ref{eq:rtw:warmaxino}) and~(\ref{eq:relaxinobbnone}). The
potentially most stringent bound from hadronic showers would require
$\maxino\gsim 360\mev$, but it is very sensitive to
$\mchi$ and disappears for $\mchi>150\gev$ (compare Fig.~4). 

At larger bino mass, the lower bound on $\maxino$ also increases,
following Eq.~(\ref{abunaavsabundchi:eq}) and because $\abundchi$
decreases with $\mchi$. Other bounds do not give any additional
constraints. In the higgsino case, one typically has $\abundchi\ll 1$
(or, more properly, the higgsino number density at freeze-out is very
much smaller than that of the bino with the same mass) owing mostly to
co-annihilation, as mentioned at the end of the last section. A heavy
higgsino with $\mchi\gsim500\gev$ nevertheless remains an option for a
LOSP. In this case, however, additional model-dependent (dimension-4)
operators will contribute to the TP mechanism, and the upper curve in
Fig.~5 will probably be moved upwards. Overall we find the bino case
to be much more natural and robust.

%%%%%%%%%%%%%%%%%%%%%%%%%%%%%%%%%%%%%%%%%%%%%%%%%%%%%%%%%%%%%%%%%%%%%%%%%%%
\section{Implications and Conclusions}
\label{sec:implications}

The intriguing possibility that the axino is the LSP and the dark
matter WIMP possesses a number of very distinct features. This makes
this case very different from those of both the neutralino and the
gravitino. In particular, the axino can be a cold DM WIMP for a rather
wide range of masses in the $\mev$ to $\gev$ range and for relatively
low reheating temperatures $\treh\lsim5\times10^4\gev$. As $\treh$
increases, thermal production of axinos starts dominating over
non-thermal production and the axino typically becomes a warm DM relic
with a mass broadly in the $\kev$ range.  In contrast, the neutralino
is typically a cold DM WIMP (although see Ref.~\cite{hisano00}).

Low reheating temperatures would favor baryogenesis at the electroweak
scale. It would also alleviate the nagging ``gravitino
problem''. If additionally it is the axino that is the LSP and the
gravitino is the NLSP, the gravitino problem is resolved altogether
for both low and high $\treh$. 

Phenomenologically, one faces a well-justified possibility that the
bound $\abundchi<1$, which is often imposed in constraining a SUSY
parameter space, may be readily avoided. In fact, the range
$\abundchi\gg 1$ (and with it typically large masses of superpartners)
would now be favored if the axino is to be a dominant component of DM
in the Universe.  Furthermore, the lightest ordinary superpartner
could either be neutral or charged but would appear stable in collider
searches.

The axino, with its exceedingly tiny coupling to other matter, will be
a real challenge to experimentalists. It is much more plausible that a
supersymmetric particle and the axion will be found first. Unless the
neutralino (or some other WIMP) is detected in DM searches, the axino
will remain an attractive and robust candidate for solving the
outstanding puzzle of the nature of dark matter in the Universe.

%
% ACKNOWLEDGMENTS
%

%\acknowledgements 
%\acknowledgments 
\section*{Acknowledgements}

JEK is supported in part by the BK21 program of Ministry of Education,
Korea Research Foundation Grant No. KRF-2000-015-DP0072, CTP Research
Fund of Seoul National University, and by the Center for High Energy
Physics (CHEP), Kyungpook National University.  LR would like to thank
A.~Dolgov, J.~March-Russell, T.~Moroi, H.P.~Nilles, and A.~Riotto for
interesting comments.  LC would like to thank T.~Asaka,
W.~Buchm\"uller, F. Madricardo, D.J.~Miller and G.~Moortgat-Pick for
useful discussions.  LC, HBK and LR would like to acknowledge the kind
hospitality and support of KIAS (Korea Institute for Advanced Study)
where part of the project has been done.

%
% REFERENCES
%


\begin{thebibliography}{99}

\def\apj#1#2#3{{\it Astrophys. J. }{\bf #1}, (#2) #3}
%\def\ibid#1#2#3{{\it ibid\/} {\bf #1}, (#2) #3}
\def\ibid{{\it ibid}}

\bibitem{ckr}
L.~Covi, J.E.~Kim and L.~Roszkowski, \prl{82}{1999}{4180}.

\bibitem{pq}
R.D.~Peccei and H.R.~Quinn, \prl{38}{1977}{1440} and \prd{16}{1977}{1791}.

\bibitem{qcdanomaly:cite}
W.A.~Bardeen, S.-H.H.~Tye, \plb{74}{1978}{229};
V.~Baluni, \prd{19}{1979}{2227}.

\bibitem{axionreviews:cite}
J.E.~Kim, \prep{150}{1987}{1};
H.Y.~Cheng, \prep{158}{1988}{1};
R.D.~Peccei, in {\it CP Violation},
ed. C.~Jarlskog (World Scientific Publishing Co., 1989);
M.S.~Turner, \prep{197}{1990}{67};
G.G.~Raffelt,  \prep{198}{1990}{1};
P.~Sikivie, \hepph{0002154}.

\bibitem{axiondm}
J.~Preskill, M.B.~Wise and F.~Wilczek, \plb{120}{1983}{127};
L.F.~Abbott and P.~Sikivie, \plb{120}{1983}{133};
M.~Dine and W.~Fischler, \plb{120}{1983}{137}.

\bibitem{moreearlygravitino}
J. Ellis, A.D.~Linde and D.V. Nanopoulos, \plb{118}{1982}{59}; 
M.Y.~Khlopov and A.D.~Linde, \ibid\ {\bf B 138} (1984) 265.

\bibitem{ekn84}
J.~Ellis, J.E.~Kim and D.V.~Nanopoulos, \plb{145}{1984}{181}.

\bibitem{mmy}
T.~Moroi, H.~Murayama and M.~Yamaguchi, \plb{303}{1993}{289}.

\bibitem{bbp98}
M.~Bolz, W.~Buchm\"uller and M.~Pl\"umacher, \plb{443}{1998}{209}.

\bibitem{ty00} 
F.~Takayama and M.~Yamaguchi, \plb{485}{2000}{388}.

\bibitem{wil}
M.~Bolz, A.~Brandenburg and W.~Buchm\"uller, \hepph{0012052}.

\bibitem{tw}
K.~Tamvakis and D.~Wyler, \plb{112}{1982}{451}.

\bibitem{rtw} 
K.~Rajagopal, M.S.~Turner and F.~Wilczek, \npb{358}{1991}{447}.

\bibitem{ckn}
E.J.~Chun, J.E.~Kim and H.P.~Nilles, \plb{287}{1992}{123}.

\bibitem{kmn}
J.E.~Kim, A.~Masiero and D.V.~Nanopoulos, \plb{139}{1984}{346}.

\bibitem{bgm}
S.A.~Bonometto, F.~Gabbiani and A.~Masiero,
\plb{222}{433}{1989} and \prd{49}{1994}{3918}.

\bibitem{bot} 
S.~Colombi, S.~Dodelson and L.M.~Widrow, \astroph{9505029};
P.~Colin, V.~Avila-Reese and O.~Valenzuela, \apj{543}{2000}{622};
P.~Bode, J.P.~Ostriker and N.~Turok, \astroph{0010389}.

\bibitem{kt}
E.W.~Kolb and M.S.~Turner,
{\it The Early Universe} (Addison-Wesley, Redwood City, 1990).

\bibitem{kkrw}
G.L.~Kane, C.~Kolda, L.~Roszkowski and J.D.~Wells, \prd{49}{1994}{6173}.

\bibitem{fm00}
J.L.~Feng and K.T.~Matchev, \prd{63}{2001}{095003}.

\bibitem{dmr00}
R.~Derm{\'\i}sek, A.~Mafi and S.~Raby, \prd{63}{2001}{035001}.

\bibitem{grt}
R.~Kallosh, L.~Kofman, A.~Linde and A.~Van Proeyen, \prd{61}{2000}{103503};
G.F.~Giudice, I.~Tkachev and A.~Riotto, \jhep{9908}{1999}{009}.

\bibitem{nps01}
H.P.~Nilles, M.~Peloso and L.~Sorbo, hep-ph/0102264 and
\jhep{0104}{2001}{004}.

\bibitem{grt01}
G.F.~Giudice, A.~Riotto and I.~Tkachev, hep-ph/0103248.

\bibitem{ay00}
T.~Asaka and T.~Yanagida, \plb{494}{297}{2000}.

\bibitem{ck} 
S.~Chang and H.B.~Kim, \prl{77}{1996}{591}.

\bibitem{kim91}
J.E.~Kim, \prl{67}{1991}{3465}; D.H.~Lyth, \prd{48}{1993}{4523}.

\bibitem{gr}
H.~Georgi and L.~Randall, \npb{276}{1986}{241}.

\bibitem{ssaxion}
E.~Witten, \plb{149}{1984}{351} and \ibid\ {\bf B
    153} (1985) 243; K.~Choi and J.E.~Kim, \plb{154}{1985}{393} and
  \ibid\ {\bf B 165} (1985) 71.

\bibitem{compaxion}
J.E.~Kim, \prd{31}{1985}{1733};
K.~Choi and J.E.~Kim, \prd{32}{1985}{1828}.

\bibitem{ksvz}
J.E.~Kim, \prl{43}{1979}{103};
M.A.~Shifman, V.I.~Vainstein and V.I.~Zakharov, \npb{166}{1980}{4933}.

\bibitem{dfsz} 
M.~Dine, W.~Fischler and M.~Srednicki, \plb{104}{1981}{99};
A.P.~Zhitnitskii, \sjnp{31}{1980}{260}.

\bibitem{axion}
S.~Weinberg, \prl{40}{1978}{223};
F.~Wilczek, \prl{40}{1978}{279}.

\bibitem{susyaxion1} 
% References of Supersymmetric axion models
H.P.~Nilles and S.~Raby, \npb{198}{1982}{102};

\bibitem{susyaxion2} 
J.E.~Kim and H.P.~Nilles, \plb{138}{1984}{150}.

\bibitem{Goto-Yamaguchi}
T.~Goto ang M.~Yamaguchi, \plb{276}{1992}{103}.

\bibitem{cl}
E.J.~Chun and A.~Lukas, \plb{357}{1995}{43}.

\bibitem{nieves}
J.F.~Nieves, \plb{174}{1986}{411}.

\bibitem{jekim83}
J.E.~Kim, \plb{136}{1984}{378}.

\bibitem{ckl00}
E.J.~Chun, H.B.~Kim and D.H.~Lyth, \prd{62}{2000}{125001}.

\bibitem{Moxhay-Yamamoto} 
P.~Moxhay and K.~Yamamoto, \plb{151}{1985}{363}.

\bibitem{kk}
E.J.~Chun, H.B.~Kim and J.E.~Kim, \prl{72}{1994}{1956};
H.B.~Kim and J.E.~Kim, \npb{433}{1995}{421}.

\bibitem{large}
D.R.~Stump, M.~Wiest and C.P.~Yuan, \prd{54}{1996}{1936}.

\bibitem{chun}
E.J.~Chun, \plb{454}{1999}{304}.

\bibitem{chiasdm}
L.~Roszkowski, \plb{262}{1991}{59}.

\bibitem{rr93} 
R.G.~Roberts and L.~Roszkowski, \plb{309}{1993}{329}.

\bibitem{coupling}
J.E.~Kim, \prd{58}{1998}{055006}.
See, also,
D.B.~Kaplan, \npb{260}{1985}{215};
M.~Srednicki, \npb{260}{1985}{689}.

\bibitem{chkl}
K.~Choi, K.~Hwang, H.B.~Kim and T.~Lee, \plb{467}{1999}{211}.

\bibitem{by91}
E.~Braaten and T.C.~Yuan, \prl{66}{1991}{2183}.

\bibitem{gkr00}
G.F.~Giudice, E.W.~Kolb and A.~Riotto, \hepph{0005123}.

\bibitem{muterm}
E.J.~Chun, J.E.~Kim and H.P.~Nilles, \npb{370}{1992}{105};
J.E.~Kim and B.~Kyae, \plb{500}{2001}{313}.

\bibitem{martinaxino}
S.P.~Martin, \prd{62}{2000}{095008}.

\bibitem{bbnbound} 
J.~Ellis \etal, \npb{373}{1992}{399}.

\bibitem{dkt}
D.A.~Dicus, E.W.~Kolb and V.L.~Teplitz, \apj{221}{1979}{327}.

\bibitem{hisano00}
W.B.~Lin, D.H.~Huang, X.~Zhang and R.~Brandenberger, \prl{86}{2001}{954};
J.~Hisano, K.~Kohri and M.M.~Nojiri, \hepph{0011216}.

\end{thebibliography}
\end{document}